\ifx\mnmacrosloaded\undefined 
%
%
%
%

\catcode `\@=11 

\def\@version{1.6}
\def\@verdate{18th September 1995}

%
%


\newif\ifprod@font

\ifx\@typeface\undefined
  \def\@typeface{Comp. Modern}\prod@fontfalse
\else
  \prod@fonttrue 
\fi

\def\newfam{\alloc@8\fam\chardef\sixt@@n} 

\ifprod@font
\font\fiverm=mtr10 at 5pt
\font\fivebf=mtbx10 at 5pt
\font\fiveit=mtti10 at 5pt
\font\fivesl=mtsl10 at 5pt
\font\fivett=cmtt8 at 5pt     \hyphenchar\fivett=-1
\font\fivecsc=mtcsc10 at 5pt
\font\fivesf=mtss10 at 5pt
\font\fivei=mtmi10 at 5pt      \skewchar\fivei='177
\font\fivesy=mtsy10 at 5pt     \skewchar\fivesy='60

\font\sixrm=mtr10 at 6pt
\font\sixbf=mtbx10 at 6pt
\font\sixit=mtti10 at 6pt
\font\sixsl=mtsl10 at 6pt
\font\sixtt=cmtt8 at 6pt      \hyphenchar\sixtt=-1
\font\sixcsc=mtcsc10 at 6pt
\font\sixsf=mtss10 at 6pt
\font\sixi=mtmi10 at 6pt       \skewchar\sixi='177
\font\sixsy=mtsy10 at 6pt      \skewchar\sixsy='60

\font\sevenrm=mtr10 at 7pt
\font\sevenbf=mtbx10 at 7pt
\font\sevenit=mtti10 at 7pt
\font\sevensl=mtsl10 at 7pt
\font\seventt=cmtt8 at 7pt     \hyphenchar\seventt=-1
\font\sevencsc=mtcsc10 at 7pt
\font\sevensf=mtss10 at 7pt
\font\seveni=mtmi10 at 7pt      \skewchar\seveni='177
\font\sevensy=mtsy10 at 7pt     \skewchar\sevensy='60

\font\eightrm=mtr10 at 8pt
\font\eightbf=mtbx10 at 8pt
\font\eightit=mtti10 at 8pt
\font\eighti=mtmi10 at 8pt      \skewchar\eighti='177
\font\eightsy=mtsy10 at 8pt     \skewchar\eightsy='60
\font\eightsl=mtsl10 at 8pt
\font\eighttt=cmtt8             \hyphenchar\eighttt=-1
\font\eightcsc=mtcsc10 at 8pt
\font\eightsf=mtss10 at 8pt

\font\ninerm=mtr10 at 9pt
\font\ninebf=mtbx10 at 9pt
\font\nineit=mtti10 at 9pt
\font\ninei=mtmi10 at 9pt      \skewchar\ninei='177
\font\ninesy=mtsy10 at 9pt     \skewchar\ninesy='60
\font\ninesl=mtsl10 at 9pt
\font\ninett=cmtt9             \hyphenchar\ninett=-1
\font\ninecsc=mtcsc10 at 9pt
\font\ninesf=mtss10 at 9pt

\font\tenrm=mtr10
\font\tenbf=mtbx10
\font\tenit=mtti10
\font\teni=mtmi10		\skewchar\teni='177
\font\tensy=mtsy10		\skewchar\tensy='60
\font\tenex=cmex10
\font\tensl=mtsl10
\font\tentt=cmtt10		\hyphenchar\tentt=-1
\font\tencsc=mtcsc10
\font\tensf=mtss10

\font\elevenrm=mtr10 at 11pt
\font\elevenbf=mtbx10 at 11pt
\font\elevenit=mtti10 at 11pt
\font\eleveni=mtmi10 at 11pt      \skewchar\eleveni='177
\font\elevensy=mtsy10 at 11pt     \skewchar\elevensy='60
\font\elevensl=mtsl10 at 11pt
\font\eleventt=cmtt10 at 11pt     \hyphenchar\eleventt=-1
\font\elevencsc=mtcsc10 at 11pt
\font\elevensf=mtss10 at 11pt

\font\twelverm=mtr10 at 12pt
\font\twelvebf=mtbx10 at 12pt
\font\twelveit=mtti10 at 12pt
\font\twelvesl=mtsl10 at 12pt
\font\twelvett=cmtt12             \hyphenchar\twelvett=-1
\font\twelvecsc=mtcsc10 at 12pt
\font\twelvesf=mtss10 at 12pt
\font\twelvei=mtmi10 at 12pt      \skewchar\twelvei='177
\font\twelvesy=mtsy10 at 12pt     \skewchar\twelvesy='60

\font\fourteenrm=mtr10 at 14pt
\font\fourteenbf=mtbx10 at 14pt
\font\fourteenit=mtti10 at 14pt
\font\fourteeni=mtmi10 at 14pt      \skewchar\fourteeni='177
\font\fourteensy=mtsy10 at 14pt     \skewchar\fourteensy='60
\font\fourteensl=mtsl10 at 14pt
\font\fourteentt=cmtt12 at 14pt     \hyphenchar\fourteentt=-1
\font\fourteencsc=mtcsc10 at 14pt
\font\fourteensf=mtss10 at 14pt

\font\seventeenrm=mtr10 at 17pt
\font\seventeenbf=mtbx10 at 17pt
\font\seventeenit=mtti10 at 17pt
\font\seventeeni=mtmi10 at 17pt      \skewchar\seventeeni='177
\font\seventeensy=mtsy10 at 17pt     \skewchar\seventeensy='60
\font\seventeensl=mtsl10 at 17pt
\font\seventeentt=cmtt12 at 17pt     \hyphenchar\seventeentt=-1
\font\seventeencsc=mtcsc10 at 17pt
\font\seventeensf=mtss10 at 17pt
\else
\font\fiverm=cmr5
\font\fivei=cmmi5             \skewchar\fivei='177
\font\fivesy=cmsy5            \skewchar\fivesy='60
\font\fivebf=cmbx5

\font\sixrm=cmr6
\font\sixi=cmmi6             \skewchar\sixi='177
\font\sixsy=cmsy6            \skewchar\sixsy='60
\font\sixbf=cmbx6

\font\sevenrm=cmr7
\font\sevenit=cmti7
\font\seveni=cmmi7             \skewchar\seveni='177
\font\sevensy=cmsy7            \skewchar\sevensy='60
\font\sevenbf=cmbx7

\font\eightrm=cmr8
\font\eightbf=cmbx8
\font\eightit=cmti8
\font\eighti=cmmi8			\skewchar\eighti='177
\font\eightsy=cmsy8			\skewchar\eightsy='60
\font\eightsl=cmsl8
\font\eighttt=cmtt8			\hyphenchar\eighttt=-1
\font\eightcsc=cmcsc10 at 8pt
\font\eightsf=cmss8

\font\ninerm=cmr9
\font\ninebf=cmbx9
\font\nineit=cmti9
\font\ninei=cmmi9			\skewchar\ninei='177
\font\ninesy=cmsy9			\skewchar\ninesy='60
\font\ninesl=cmsl9
\font\ninett=cmtt9			\hyphenchar\ninett=-1
\font\ninecsc=cmcsc10 at 9pt
\font\ninesf=cmss9

\font\tenrm=cmr10
\font\tenbf=cmbx10
\font\tenit=cmti10
\font\teni=cmmi10		\skewchar\teni='177
\font\tensy=cmsy10		\skewchar\tensy='60
\font\tenex=cmex10
\font\tensl=cmsl10
\font\tentt=cmtt10		\hyphenchar\tentt=-1
\font\tencsc=cmcsc10
\font\tensf=cmss10

\font\elevenrm=cmr10 scaled \magstephalf
\font\elevenbf=cmbx10 scaled \magstephalf
\font\elevenit=cmti10 scaled \magstephalf
\font\eleveni=cmmi10 scaled \magstephalf	\skewchar\eleveni='177
\font\elevensy=cmsy10 scaled \magstephalf	\skewchar\elevensy='60
\font\elevensl=cmsl10 scaled \magstephalf
\font\eleventt=cmtt10 scaled \magstephalf	\hyphenchar\eleventt=-1
\font\elevencsc=cmcsc10 scaled \magstephalf
\font\elevensf=cmss10 scaled \magstephalf

\font\twelverm=cmr10 scaled \magstep1
\font\twelvebf=cmbx10 scaled \magstep1
\font\twelvei=cmmi10 scaled \magstep1      \skewchar\twelvei='177
\font\twelvesy=cmsy10 scaled \magstep1     \skewchar\twelvesy='60

\font\fourteenrm=cmr10 scaled \magstep2
\font\fourteenbf=cmbx10 scaled \magstep2
\font\fourteenit=cmti10 scaled \magstep2
\font\fourteeni=cmmi10 scaled \magstep2		\skewchar\fourteeni='177
\font\fourteensy=cmsy10 scaled \magstep2	\skewchar\fourteensy='60
\font\fourteensl=cmsl10 scaled \magstep2
\font\fourteentt=cmtt10 scaled \magstep2	\hyphenchar\fourteentt=-1
\font\fourteencsc=cmcsc10 scaled \magstep2
\font\fourteensf=cmss10 scaled \magstep2

\font\seventeenrm=cmr10 scaled \magstep3
\font\seventeenbf=cmbx10 scaled \magstep3
\font\seventeenit=cmti10 scaled \magstep3
\font\seventeeni=cmmi10 scaled \magstep3	\skewchar\seventeeni='177
\font\seventeensy=cmsy10 scaled \magstep3	\skewchar\seventeensy='60
\font\seventeensl=cmsl10 scaled \magstep3
\font\seventeentt=cmtt10 scaled \magstep3	\hyphenchar\seventeentt=-1
\font\seventeencsc=cmcsc10 scaled \magstep3
\font\seventeensf=cmss10 scaled \magstep3
\fi

\def\hexnumber#1{\ifcase#1 0\or1\or2\or3\or4\or5\or6\or7\or8\or9\or
  A\or B\or C\or D\or E\or F\fi}

\def\makestrut{%
  \setbox\strutbox=\hbox{%
    \vrule height.7\baselineskip depth.3\baselineskip width \z@}%
}

\def\baselinestretch{1}
\newskip\tmp@bls

\def\b@ls#1{
  \tmp@bls=#1\relax
  \baselineskip=#1\relax\makestrut
  \normalbaselineskip=\baselinestretch\tmp@bls
  \normalbaselines
}

\def\nostb@ls#1{
  \normalbaselineskip=#1\relax
  \normalbaselines
  \makestrut
}

%

\newfam\scfam  
\newfam\sffam  

\def\mit{\fam\@ne}
\def\cal{\fam\tw@}
\def\em{\ifdim\fontdimen1\font>\z@ \rm\else\it\fi}

\textfont3=\tenex
\scriptfont3=\tenex
\scriptscriptfont3=\tenex

\setbox0=\hbox{\tenex B} \p@renwd=\wd0 

\def\eightpoint{
  \def\rm{\fam0\eightrm}%
  \textfont0=\eightrm \scriptfont0=\sixrm \scriptscriptfont0=\fiverm%
  \textfont1=\eighti  \scriptfont1=\sixi  \scriptscriptfont1=\fivei%
  \textfont2=\eightsy \scriptfont2=\sixsy \scriptscriptfont2=\fivesy%
  \textfont\itfam=\eightit\def\it{\fam\itfam\eightit}%
  \ifprod@font
    \scriptfont\itfam=\sixit
      \scriptscriptfont\itfam=\fiveit
  \else
    \scriptfont\itfam=\eightit
      \scriptscriptfont\itfam=\eightit
  \fi
  \textfont\bffam=\eightbf%
    \scriptfont\bffam=\sixbf%
      \scriptscriptfont\bffam=\fivebf%
  \def\bf{\fam\bffam\eightbf}%
  \textfont\slfam=\eightsl\def\sl{\fam\slfam\eightsl}%
  \ifprod@font
    \scriptfont\slfam=\sixsl
      \scriptscriptfont\slfam=\fivesl
  \else
    \scriptfont\slfam=\eightsl
      \scriptscriptfont\slfam=\eightsl
  \fi
  \textfont\ttfam=\eighttt\def\tt{\fam\ttfam\eighttt}%
  \ifprod@font
    \scriptfont\ttfam=\sixtt
      \scriptscriptfont\ttfam=\fivett
  \else
    \scriptfont\ttfam=\eighttt
      \scriptscriptfont\ttfam=\eighttt
  \fi
  \textfont\scfam=\eightcsc\def\sc{\fam\scfam\eightcsc}%
  \ifprod@font
    \scriptfont\scfam=\sixcsc
      \scriptscriptfont\scfam=\fivecsc
  \else
    \scriptfont\scfam=\eightcsc
      \scriptscriptfont\scfam=\eightcsc
  \fi
  \textfont\sffam=\eightsf\def\sf{\fam\sffam\eightsf}%
  \ifprod@font
    \scriptfont\sffam=\sixsf
      \scriptscriptfont\sffam=\fivesf
  \else
    \scriptfont\sffam=\eightsf
      \scriptscriptfont\sffam=\eightsf
  \fi
  \def\oldstyle{\fam\@ne\eighti}%
  \b@ls{10pt}\rm\@viiipt%
}
\def\@viiipt{}

\def\ninepoint{
  \def\rm{\fam0\ninerm}%
  \textfont0=\ninerm \scriptfont0=\sixrm \scriptscriptfont0=\fiverm%
  \textfont1=\ninei  \scriptfont1=\sixi  \scriptscriptfont1=\fivei%
  \textfont2=\ninesy \scriptfont2=\sixsy \scriptscriptfont2=\fivesy%
  \textfont\itfam=\nineit\def\it{\fam\itfam\nineit}%
  \ifprod@font
    \scriptfont\itfam=\sixit
      \scriptscriptfont\itfam=\fiveit
  \else
    \scriptfont\itfam=\nineit
      \scriptscriptfont\itfam=\nineit
  \fi
  \textfont\bffam=\ninebf%
    \scriptfont\bffam=\sixbf%
      \scriptscriptfont\bffam=\fivebf%
  \def\bf{\fam\bffam\ninebf}%
  \textfont\slfam=\ninesl\def\sl{\fam\slfam\ninesl}%
  \ifprod@font
    \scriptfont\slfam=\sixsl
      \scriptscriptfont\slfam=\fivesl
  \else
    \scriptfont\slfam=\ninesl
      \scriptscriptfont\slfam=\ninesl
  \fi
  \textfont\ttfam=\ninett\def\tt{\fam\ttfam\ninett}%
  \ifprod@font
    \scriptfont\ttfam=\sixtt
      \scriptscriptfont\ttfam=\fivett
  \else
    \scriptfont\ttfam=\ninett
      \scriptscriptfont\ttfam=\ninett
  \fi
  \textfont\scfam=\ninecsc\def\sc{\fam\scfam\ninecsc}%
  \ifprod@font
    \scriptfont\scfam=\sixcsc
      \scriptscriptfont\scfam=\fivecsc
  \else
    \scriptfont\scfam=\ninecsc
      \scriptscriptfont\scfam=\ninecsc
  \fi
  \textfont\sffam=\ninesf\def\sf{\fam\sffam\ninesf}%
  \ifprod@font
    \scriptfont\sffam=\sixsf
      \scriptscriptfont\sffam=\fivesf
  \else
    \scriptfont\sffam=\ninesf
      \scriptscriptfont\sffam=\ninesf
  \fi
  \def\oldstyle{\fam\@ne\ninei}%
  \b@ls{\TextLeading plus \Feathering}\rm\@ixpt%
}
\def\@ixpt{}

\def\tenpoint{
  \def\rm{\fam0\tenrm}%
  \textfont0=\tenrm \scriptfont0=\sevenrm \scriptscriptfont0=\fiverm%
  \textfont1=\teni  \scriptfont1=\seveni  \scriptscriptfont1=\fivei%
  \textfont2=\tensy \scriptfont2=\sevensy \scriptscriptfont2=\fivesy%
  \textfont\itfam=\tenit\def\it{\fam\itfam\tenit}%
  \ifprod@font
    \scriptfont\itfam=\sevenit
      \scriptscriptfont\itfam=\fiveit
  \else
    \scriptfont\itfam=\tenit
      \scriptscriptfont\itfam=\tenit
  \fi
  \textfont\bffam=\tenbf%
    \scriptfont\bffam=\sevenbf%
      \scriptscriptfont\bffam=\fivebf%
  \def\bf{\fam\bffam\tenbf}%
  \textfont\slfam=\tensl\def\sl{\fam\slfam\tensl}%
  \ifprod@font
    \scriptfont\slfam=\sevensl
      \scriptscriptfont\slfam=\fivesl
  \else
    \scriptfont\slfam=\tensl
      \scriptscriptfont\slfam=\tensl
  \fi
  \textfont\ttfam=\tentt\def\tt{\fam\ttfam\tentt}%
  \ifprod@font
    \scriptfont\ttfam=\seventt
      \scriptscriptfont\ttfam=\fivett
  \else
    \scriptfont\ttfam=\tentt
      \scriptscriptfont\ttfam=\tentt
  \fi
  \textfont\scfam=\tencsc\def\sc{\fam\scfam\tencsc}%
  \ifprod@font
    \scriptfont\scfam=\sevencsc
      \scriptscriptfont\scfam=\fivecsc
  \else
    \scriptfont\scfam=\tencsc
      \scriptscriptfont\scfam=\tencsc
  \fi
  \textfont\sffam=\tensf\def\sf{\fam\sffam\tensf}%
  \ifprod@font
    \scriptfont\sffam=\sevensf
      \scriptscriptfont\sffam=\fivesf
  \else
    \scriptfont\sffam=\tensf
      \scriptscriptfont\sffam=\tensf
  \fi
  \def\oldstyle{\fam\@ne\teni}%
  \b@ls{11pt}\rm\@xpt%
}
\def\@xpt{}

\def\elevenpoint{
  \def\rm{\fam0\elevenrm}%
  \textfont0=\elevenrm \scriptfont0=\eightrm \scriptscriptfont0=\sixrm%
  \textfont1=\eleveni  \scriptfont1=\eighti  \scriptscriptfont1=\sixi%
  \textfont2=\elevensy \scriptfont2=\eightsy \scriptscriptfont2=\sixsy%
  \textfont\itfam=\elevenit\def\it{\fam\itfam\elevenit}%
  \ifprod@font
    \scriptfont\itfam=\eightit
      \scriptscriptfont\itfam=\sixit
  \else
    \scriptfont\itfam=\elevenit
      \scriptscriptfont\itfam=\elevenit
  \fi
  \textfont\bffam=\elevenbf%
    \scriptfont\bffam=\eightbf%
      \scriptscriptfont\bffam=\sixbf%
  \def\bf{\fam\bffam\elevenbf}%
  \textfont\slfam=\elevensl\def\sl{\fam\slfam\elevensl}%
  \ifprod@font
    \scriptfont\slfam=\eightsl
      \scriptscriptfont\slfam=\sixsl
  \else
    \scriptfont\slfam=\elevensl
      \scriptscriptfont\slfam=\elevensl
  \fi
  \textfont\ttfam=\eleventt\def\tt{\fam\ttfam\eleventt}%
  \ifprod@font
    \scriptfont\ttfam=\eighttt
      \scriptscriptfont\ttfam=\sixtt
  \else
    \scriptfont\ttfam=\eleventt
      \scriptscriptfont\ttfam=\eleventt
  \fi
  \textfont\scfam=\elevencsc\def\sc{\fam\scfam\elevencsc}%
  \ifprod@font
    \scriptfont\scfam=\eightcsc
      \scriptscriptfont\scfam=\sixcsc
  \else
    \scriptfont\scfam=\elevencsc
      \scriptscriptfont\scfam=\elevencsc
  \fi
  \textfont\sffam=\elevensf\def\sf{\fam\sffam\elevensf}%
  \ifprod@font
    \scriptfont\sffam=\eightsf
      \scriptscriptfont\sffam=\sixsf
  \else
    \scriptfont\sffam=\elevensf
      \scriptscriptfont\sffam=\elevensf
  \fi
  \def\oldstyle{\fam\@ne\eleveni}%
  \b@ls{13pt}\rm\@xipt%
}
\def\@xipt{}

\def\fourteenpoint{
  \def\rm{\fam0\fourteenrm}%
  \textfont0\fourteenrm  \scriptfont0\tenrm  \scriptscriptfont0\sevenrm%
  \textfont1\fourteeni   \scriptfont1\teni   \scriptscriptfont1\seveni%
  \textfont2\fourteensy  \scriptfont2\tensy  \scriptscriptfont2\sevensy%
  \textfont\itfam=\fourteenit\def\it{\fam\itfam\fourteenit}%
  \ifprod@font
    \scriptfont\itfam=\tenit
      \scriptscriptfont\itfam=\sevenit
  \else
    \scriptfont\itfam=\fourteenit
      \scriptscriptfont\itfam=\fourteenit
  \fi
  \textfont\bffam=\fourteenbf%
    \scriptfont\bffam=\tenbf%
      \scriptscriptfont\bffam=\sevenbf%
  \def\bf{\fam\bffam\fourteenbf}%
  \textfont\slfam=\fourteensl\def\sl{\fam\slfam\fourteensl}%
  \ifprod@font
    \scriptfont\slfam=\tensl
      \scriptscriptfont\slfam=\sevensl
  \else
    \scriptfont\slfam=\fourteensl
      \scriptscriptfont\slfam=\fourteensl
  \fi
  \textfont\ttfam=\fourteentt\def\tt{\fam\ttfam\fourteentt}%
  \ifprod@font
    \scriptfont\ttfam=\tentt
      \scriptscriptfont\ttfam=\seventt
  \else
    \scriptfont\ttfam=\fourteentt
      \scriptscriptfont\ttfam=\fourteentt
  \fi
  \textfont\scfam=\fourteencsc\def\sc{\fam\scfam\fourteencsc}%
  \ifprod@font
    \scriptfont\scfam=\tencsc
      \scriptscriptfont\scfam=\sevencsc
  \else
    \scriptfont\scfam=\fourteencsc
      \scriptscriptfont\scfam=\fourteencsc
  \fi
  \textfont\sffam=\fourteensf\def\sf{\fam\sffam\fourteensf}%
  \ifprod@font
    \scriptfont\sffam=\tensf
      \scriptscriptfont\sffam=\sevensf
  \else
    \scriptfont\sffam=\fourteensf
      \scriptscriptfont\sffam=\fourteensf
  \fi
  \def\oldstyle{\fam\@ne\fourteeni}%
  \b@ls{17pt}\rm\@xivpt%
}
\def\@xivpt{}

\def\seventeenpoint{
  \def\rm{\fam0\seventeenrm}%
  \textfont0\seventeenrm  \scriptfont0\twelverm  \scriptscriptfont0\tenrm%
  \textfont1\seventeeni   \scriptfont1\twelvei   \scriptscriptfont1\teni%
  \textfont2\seventeensy  \scriptfont2\twelvesy  \scriptscriptfont2\tensy%
  \textfont\itfam=\seventeenit\def\it{\fam\itfam\seventeenit}%
  \ifprod@font
    \scriptfont\itfam=\twelveit
      \scriptscriptfont\itfam=\tenit
  \else
    \scriptfont\itfam=\seventeenit
      \scriptscriptfont\itfam=\seventeenit
  \fi
  \textfont\bffam=\seventeenbf%
    \scriptfont\bffam=\twelvebf%
      \scriptscriptfont\bffam=\tenbf%
  \def\bf{\fam\bffam\seventeenbf}%
  \textfont\slfam=\seventeensl\def\sl{\fam\slfam\seventeensl}%
  \ifprod@font
    \scriptfont\slfam=\twelvesl
      \scriptscriptfont\slfam=\tensl
  \else
    \scriptfont\slfam=\seventeensl
      \scriptscriptfont\slfam=\seventeensl
  \fi
  \textfont\ttfam=\seventeentt\def\tt{\fam\ttfam\seventeentt}%
  \ifprod@font
    \scriptfont\ttfam=\twelvett
      \scriptscriptfont\ttfam=\tentt
  \else
    \scriptfont\ttfam=\seventeentt
      \scriptscriptfont\ttfam=\seventeentt
  \fi
  \textfont\scfam=\seventeencsc\def\sc{\fam\scfam\seventeencsc}%
  \ifprod@font
    \scriptfont\scfam=\twelvecsc
      \scriptscriptfont\scfam=\tencsc
  \else
    \scriptfont\scfam=\seventeencsc
      \scriptscriptfont\scfam=\seventeencsc
  \fi
  \textfont\sffam=\seventeensf\def\sf{\fam\sffam\seventeensf}%
  \ifprod@font
    \scriptfont\sffam=\twelvesf
      \scriptscriptfont\sffam=\tensf
  \else
    \scriptfont\sffam=\seventeensf
      \scriptscriptfont\sffam=\seventeensf
  \fi
  \def\oldstyle{\fam\@ne\seventeeni}%
  \b@ls{20pt}\rm\@xviipt%
}
\def\@xviipt{}

\lineskip=1pt      \normallineskip=\lineskip
\lineskiplimit=\z@ \normallineskiplimit=\lineskiplimit


\def\loadboldmathnames{%
  \def\balpha{{\bmath{\alpha}}}%
  \def\bbeta{{\bmath{\beta}}}%
  \def\bgamma{{\bmath{\gamma}}}%
  \def\bdelta{{\bmath{\delta}}}%
  \def\bepsilon{{\bmath{\epsilon}}}%
  \def\bzeta{{\bmath{\zeta}}}%
  \def\boldeta{{\bmath{\eta}}}%
  \def\btheta{{\bmath{\theta}}}%
  \def\biota{{\bmath{\iota}}}%
  \def\bkappa{{\bmath{\kappa}}}%
  \def\blambda{{\bmath{\lambda}}}%
  \def\bmu{{\bmath{\mu}}}%
  \def\bnu{{\bmath{\nu}}}%
  \def\bxi{{\bmath{\xi}}}%
  \def\bpi{{\bmath{\pi}}}%
  \def\brho{{\bmath{\rho}}}%
  \def\bsigma{{\bmath{\sigma}}}%
  \def\btau{{\bmath{\tau}}}%
  \def\bupsilon{{\bmath{\upsilon}}}%
  \def\bphi{{\bmath{\phi}}}%
  \def\bchi{{\bmath{\chi}}}%
  \def\bpsi{{\bmath{\psi}}}%
  \def\bomega{{\bmath{\omega}}}%
  \def\bvarepsilon{{\bmath{\varepsilon}}}%
  \def\bvartheta{{\bmath{\vartheta}}}%
  \def\bvarpi{{\bmath{\varpi}}}%
  \def\bvarrho{{\bmath{\varrho}}}%
  \def\bvarsigma{{\bmath{\varsigma}}}%
  \def\bvarphi{{\bmath{\varphi}}}%
  \def\baleph{{\bmath{\aleph}}}%
  \def\bimath{{\bmath{\imath}}}%
  \def\bjmath{{\bmath{\jmath}}}%
  \def\bell{{\bmath{\ell}}}%
  \def\bwp{{\bmath{\wp}}}%
  \def\bRe{{\bmath{\Re}}}%
  \def\bIm{{\bmath{\Im}}}%
  \def\bpartial{{\bmath{\partial}}}%
  \def\binfty{{\bmath{\infty}}}%
  \def\bprime{{\bmath{\prime}}}%
  \def\bemptyset{{\bmath{\emptyset}}}%
  \def\bnabla{{\bmath{\nabla}}}%
  \def\btop{{\bmath{\top}}}%
  \def\bbot{{\bmath{\bot}}}%
  \def\btriangle{{\bmath{\triangle}}}%
  \def\bforall{{\bmath{\forall}}}%
  \def\bexists{{\bmath{\exists}}}%
  \def\bneg{{\bmath{\neg}}}%
  \def\bflat{{\bmath{\flat}}}%
  \def\bnatural{{\bmath{\natural}}}%
  \def\bsharp{{\bmath{\sharp}}}%
  \def\bclubsuit{{\bmath{\clubsuit}}}%
  \def\bdiamondsuit{{\bmath{\diamondsuit}}}%
  \def\bheartsuit{{\bmath{\heartsuit}}}%
  \def\bspadesuit{{\bmath{\spadesuit}}}%
  \def\bsmallint{{\bmath{\smallint}}}%
  \def\btriangleleft{{\bmath{\triangleleft}}}%
  \def\btriangleright{{\bmath{\triangleright}}}%
  \def\bbigtriangleup{{\bmath{\bigtriangleup}}}%
  \def\bbigtriangledown{{\bmath{\bigtriangledown}}}%
  \def\bwedge{{\bmath{\wedge}}}%
  \def\bvee{{\bmath{\vee}}}%
  \def\bcap{{\bmath{\cap}}}%
  \def\bcup{{\bmath{\cup}}}%
  \def\bddagger{{\bmath{\ddagger}}}%
  \def\bdagger{{\bmath{\dagger}}}%
  \def\bsqcap{{\bmath{\sqcap}}}%
  \def\bsqcup{{\bmath{\sqcup}}}%
  \def\buplus{{\bmath{\uplus}}}%
  \def\bamalg{{\bmath{\amalg}}}%
  \def\bdiamond{{\bmath{\diamond}}}%
  \def\bbullet{{\bmath{\bullet}}}%
  \def\bwr{{\bmath{\wr}}}%
  \def\bdiv{{\bmath{\div}}}%
  \def\bodot{{\bmath{\odot}}}%
  \def\boslash{{\bmath{\oslash}}}%
  \def\botimes{{\bmath{\otimes}}}%
  \def\bominus{{\bmath{\ominus}}}%
  \def\boplus{{\bmath{\oplus}}}%
  \def\bmp{{\bmath{\mp}}}%
  \def\bpm{{\bmath{\pm}}}%
  \def\bcirc{{\bmath{\circ}}}%
  \def\bbigcirc{{\bmath{\bigcirc}}}%
  \def\bsetminus{{\bmath{\setminus}}}%
  \def\bcdot{{\bmath{\cdot}}}%
  \def\bast{{\bmath{\ast}}}%
  \def\btimes{{\bmath{\times}}}%
  \def\bstar{{\bmath{\star}}}%
  \def\bpropto{{\bmath{\propto}}}%
  \def\bsqsubseteq{{\bmath{\sqsubseteq}}}%
  \def\bsqsupseteq{{\bmath{\sqsupseteq}}}%
  \def\bparallel{{\bmath{\parallel}}}%
  \def\bmid{{\bmath{\mid}}}%
  \def\bdashv{{\bmath{\dashv}}}%
  \def\bvdash{{\bmath{\vdash}}}%
  \def\bnearrow{{\bmath{\nearrow}}}%
  \def\bsearrow{{\bmath{\searrow}}}%
  \def\bnwarrow{{\bmath{\nwarrow}}}%
  \def\bswarrow{{\bmath{\swarrow}}}%
  \def\bLeftrightarrow{{\bmath{\Leftrightarrow}}}%
  \def\bLeftarrow{{\bmath{\Leftarrow}}}%
  \def\bRightarrow{{\bmath{\Rightarrow}}}%
  \def\bleq{{\bmath{\leq}}}%
  \def\bgeq{{\bmath{\geq}}}%
  \def\bsucc{{\bmath{\succ}}}%
  \def\bprec{{\bmath{\prec}}}%
  \def\bapprox{{\bmath{\approx}}}%
  \def\bsucceq{{\bmath{\succeq}}}%
  \def\bpreceq{{\bmath{\preceq}}}%
  \def\bsupset{{\bmath{\supset}}}%
  \def\bsubset{{\bmath{\subset}}}%
  \def\bsupseteq{{\bmath{\supseteq}}}%
  \def\bsubseteq{{\bmath{\subseteq}}}%
  \def\bin{{\bmath{\in}}}%
  \def\bni{{\bmath{\ni}}}%
  \def\bgg{{\bmath{\gg}}}%
  \def\bll{{\bmath{\ll}}}%
  \def\bnot{{\bmath{\not}}}%
  \def\bleftrightarrow{{\bmath{\leftrightarrow}}}%
  \def\bleftarrow{{\bmath{\leftarrow}}}%
  \def\brightarrow{{\bmath{\rightarrow}}}%
  \def\bmapstochar{{\bmath{\mapstochar}}}%
  \def\bsim{{\bmath{\sim}}}%
  \def\bsimeq{{\bmath{\simeq}}}%
  \def\bperp{{\bmath{\perp}}}%
  \def\bequiv{{\bmath{\equiv}}}%
  \def\basymp{{\bmath{\asymp}}}%
  \def\bsmile{{\bmath{\smile}}}%
  \def\bfrown{{\bmath{\frown}}}%
  \def\bleftharpoonup{{\bmath{\leftharpoonup}}}%
  \def\bleftharpoondown{{\bmath{\leftharpoondown}}}%
  \def\brightharpoonup{{\bmath{\rightharpoonup}}}%
  \def\brightharpoondown{{\bmath{\rightharpoondown}}}%
  \def\blhook{{\bmath{\lhook}}}%
  \def\brhook{{\bmath{\rhook}}}%
  \def\bldotp{{\bmath{\ldotp}}}%
  \def\bcdotp{{\bmath{\cdotp}}}%
}

\def\,{\relax\ifmmode \mskip\thinmuskip\else \thinspace\fi}
\let\protect=\relax

\long\def\@ifundefined#1#2#3{\expandafter\ifx\csname
  #1\endcsname\relax#2\else#3\fi}




\newtoks\math@groups \math@groups={}
\def\addtom@thgroup#1#2{#1\expandafter{\the#1#2}} 



\def\addtosizeh@ok#1#2#3#4{%
  \expandafter\def\csname @#1pt\endcsname{%
    \def\s@ze{#2}\def\ss@ze{#3}\def\sss@ze{#4}\the\math@groups%
  }%
}



\let\resetsizehook=\addtosizeh@ok


\ifprod@font
  \addtosizeh@ok{viii} {8} {6}  {5}
  \addtosizeh@ok{ix}   {9} {6}  {5}
  \addtosizeh@ok{x}    {10}{7}  {5}
  \addtosizeh@ok{xi}   {11}{8}  {6}
  \addtosizeh@ok{xiv}  {14}{10} {7}
  \addtosizeh@ok{xvii} {17}{12}{10}
\else
  \addtosizeh@ok{viii} {8}     {6}     {5}
  \addtosizeh@ok{ix}   {9}     {6}     {5}
  \addtosizeh@ok{x}    {10}    {7}     {5}
  \addtosizeh@ok{xi}   {10.95} {8}     {6}
  \addtosizeh@ok{xiv}  {14.4}  {10}    {7}
  \addtosizeh@ok{xvii} {17.28} {12}    {10}
\fi

\def\get@font#1#2#3{%
  \edef\fonts@ze{\romannumeral#3}
  \edef\fontn@me{\fonts@ze#1}
  \@ifundefined{\fontn@me}%
    {
     \global\expandafter\font\csname \fontn@me\endcsname=#2 at #3pt}%
    {}%
}

\def\ass@tfont#1#2{%
  \xdef\fam@name{\csname #1\endcsname}%
  \xdef\font@name{\csname #2\endcsname}%
  \let\textfont@name\font@name
  \textfont\fam@name\textfont@name
}

\def\ass@sfont#1#2{%
  \xdef\fam@name{\csname #1\endcsname}%
  \xdef\font@name{\csname #2\endcsname}%
  \let\textfont@name\font@name
  \scriptfont\fam@name\textfont@name
}

\def\ass@ssfont#1#2{%
  \xdef\fam@name{\csname #1\endcsname}%
  \xdef\font@name{\csname #2\endcsname}%
  \let\textfont@name\font@name
  \scriptscriptfont\fam@name\textfont@name
}


\def\NewSymbolFont#1#2{%
  \expandafter\ifx\csname sym#1fam\endcsname\relax 
    \expandafter\newfam\csname sym#1fam\endcsname
    \expandafter\edef\csname sym#1fam\endcsname{\the\allocationnumber}%
    \addtom@thgroup\math@groups{%
      \get@font{#1}{#2}{\s@ze}%
      \ass@tfont{sym#1fam}{\fontn@me}%
      \get@font{#1}{#2}{\ss@ze}%
      \ass@sfont{sym#1fam}{\fontn@me}%
      \get@font{#1}{#2}{\sss@ze}%
      \ass@ssfont{sym#1fam}{\fontn@me}%
    }%
  \else
    \errmessage{Family `#1' already defined}%
  \fi
}


\def\NewMathSymbol#1#2#3#4{%
  \edef\f@mly{\expandafter\hexnumber{\csname sym#3fam\endcsname}}%
  \mathchardef#1="#2\f@mly#4\relax
}


\newif\ifd@f

\def\NewMathDelimiter#1#2#3#4#5#6{%
  \d@ftrue
  \expandafter\ifx\csname sym#3fam\endcsname\relax
    \d@ffalse \errmessage{Family `#3' is not defined}%
  \fi
  \expandafter\ifx\csname sym#5fam\endcsname\relax
    \d@ffalse \errmessage{Family `#5' is not defined}%
  \fi
  \ifd@f
    \edef\f@mly{\expandafter\hexnumber{\csname sym#3fam\endcsname}}%
    \edef\f@mlytw@{\expandafter\hexnumber{\csname sym#5fam\endcsname}}%
    \xdef#1{\delimiter"#2\f@mly #4\f@mlytw@ #6\relax}%
  \fi
}


\def\setboxz@h{\setbox\z@\hbox}
\def\wdz@{\wd\z@}
\def\boxz@{\box\z@}
\def\setbox@ne{\setbox\@ne}
\def\wd@ne{\wd\@ne}

\def\math@atom#1#2{%
   \binrel@{#1}\binrel@@{#2}}
\def\binrel@#1{\setboxz@h{\thinmuskip0mu
  \medmuskip\m@ne mu\thickmuskip\@ne mu$#1\m@th$}%
 \setbox@ne\hbox{\thinmuskip0mu\medmuskip\m@ne mu\thickmuskip
  \@ne mu${}#1{}\m@th$}%
 \setbox\tw@\hbox{\hskip\wd@ne\hskip-\wdz@}}
\def\binrel@@#1{\ifdim\wd2<\z@\mathbin{#1}\else\ifdim\wd\tw@>\z@
 \mathrel{#1}\else{#1}\fi\fi}

\def\m@thit{1}

\def\set@skchar#1{\global\expandafter\skewchar
  \csname\fontn@me\endcsname=#1\relax}

\def\NewMathAlphabet#1#2#3{%
  \def\tst{#3}%
  \ifx\tst\empty\else 
    \expandafter\gdef\csname #1@sc\endcsname{}
  \fi
  \expandafter\def\csname #1\endcsname{
    \protect\csname @#1\endcsname}%
  \expandafter\def\csname @#1\endcsname##1{
    {%
    \begingroup
      \get@font{#1}{#2}{\s@ze}%
      \@ifundefined{#1@sc}{}{\set@skchar{#3}}%
      \ass@tfont{m@thit}{\fontn@me}%
      \get@font{#1}{#2}{\ss@ze}%
      \@ifundefined{#1@sc}{}{\set@skchar{#3}}%
      \ass@sfont{m@thit}{\fontn@me}%
      \get@font{#1}{#2}{\sss@ze}%
      \@ifundefined{#1@sc}{}{\set@skchar{#3}}%
      \ass@ssfont{m@thit}{\fontn@me}%
      \math@atom{##1}{%
      \mathchoice%
        {\hbox{$\m@th\displaystyle##1$}}%
        {\hbox{$\m@th\textstyle##1$}}%
        {\hbox{$\m@th\scriptstyle##1$}}%
        {\hbox{$\m@th\scriptscriptstyle##1$}}}%
    \endgroup
    }%
  }%
}


\newif\iffirstta  \firsttatrue

\def\set@hchar#1{\global\expandafter\hyphenchar
  \csname\fontn@me\endcsname=#1\relax}

\def\NewTextAlphabet#1#2#3{%
  \iffirstta
    \global\firsttafalse
    \newfam\scratchfam
    \edef\scrt@fam{\the\allocationnumber}%
  \fi
  \def\tst{#3}%
  \ifx\tst\empty\else 
    \expandafter\gdef\csname #1@hc\endcsname{}
  \fi
  \expandafter\def\csname #1\endcsname{
    \protect\csname t@#1\endcsname}%
  \long\expandafter\def\csname t@#1\endcsname##1{
    \ifmmode
      \typeout{Warning: do not use \expandafter\string\csname #1\endcsname
        \space in math mode}\fi%
    {%
      \get@font{#1}{#2}{\s@ze}\let\t@xtfnt=\fontn@me\relax
      \@ifundefined{#1@hc}{}{\set@hchar{#3}}%
      \ass@tfont{scrt@fam}{\fontn@me}%
      \get@font{#1}{#2}{\ss@ze}%
      \@ifundefined{#1@hc}{}{\set@hchar{#3}}%
      \ass@sfont{scrt@fam}{\fontn@me}%
      \get@font{#1}{#2}{\sss@ze}%
      \@ifundefined{#1@hc}{}{\set@hchar{#3}}%
      \ass@ssfont{scrt@fam}{\fontn@me}%
      \fam\scratchfam\csname\t@xtfnt\endcsname
    ##1%
    }%
  }%
  \expandafter\def\csname #1shape
    \endcsname{\protect\csname @#1shape\endcsname}%
  \expandafter\def\csname @#1shape\endcsname{
    \ifmmode
      \typeout{Warning: do not use \expandafter\string\csname
        #1shape\endcsname \space in math mode}\fi
      \get@font{#1}{#2}{\s@ze}\let\t@xtfnt=\fontn@me\relax
      \@ifundefined{#1@hc}{}{\set@hchar{#3}}%
      \ass@tfont{scrt@fam}{\fontn@me}%
      \get@font{#1}{#2}{\ss@ze}%
      \@ifundefined{#1@hc}{}{\set@hchar{#3}}%
      \ass@sfont{scrt@fam}{\fontn@me}%
      \get@font{#1}{#2}{\sss@ze}%
      \@ifundefined{#1@hc}{}{\set@hchar{#3}}%
      \ass@ssfont{scrt@fam}{\fontn@me}%
      \fam\scratchfam\csname\t@xtfnt\endcsname
  }%
}


\ifprod@font
  \def\math@itfnt{mtmib10}
  \def\math@syfnt{mtbsy10}
\else
  \def\math@itfnt{cmmib10}
  \def\math@syfnt{cmbsy10}
\fi

\def\m@thsy{2}

\def\bmath{\protect\@bmath}
\def\@bmath#1{%
  {%
  \begingroup
    \get@font{mthit}{\math@itfnt}{\s@ze}\set@skchar{'177}%
    \ass@tfont{m@thit}{\fontn@me}%
    \get@font{mthit}{\math@itfnt}{\ss@ze}\set@skchar{'177}%
    \ass@sfont{m@thit}{\fontn@me}%
    \get@font{mthit}{\math@itfnt}{\sss@ze}\set@skchar{'177}%
    \ass@ssfont{m@thit}{\fontn@me}%
    \get@font{mthsy}{\math@syfnt}{\s@ze}\set@skchar{'60}%
    \ass@tfont{m@thsy}{\fontn@me}%
    \get@font{mthsy}{\math@syfnt}{\ss@ze}\set@skchar{'60}%
    \ass@sfont{m@thsy}{\fontn@me}%
    \get@font{mthsy}{\math@syfnt}{\sss@ze}\set@skchar{'60}%
    \ass@ssfont{m@thsy}{\fontn@me}%
    \math@atom{#1}{%
    \mathchoice%
      {\hbox{$\m@th\displaystyle#1$}}%
      {\hbox{$\m@th\textstyle#1$}}%
      {\hbox{$\m@th\scriptstyle#1$}}%
      {\hbox{$\m@th\scriptscriptstyle#1$}}}%
  \endgroup
  }%
}



\def\la{\mathrel{\mathchoice {\vcenter{\offinterlineskip\halign{\hfil
$\displaystyle##$\hfil\cr<\cr\sim\cr}}}
{\vcenter{\offinterlineskip\halign{\hfil$\textstyle##$\hfil\cr
<\cr\sim\cr}}}
{\vcenter{\offinterlineskip\halign{\hfil$\scriptstyle##$\hfil\cr
<\cr\sim\cr}}}
{\vcenter{\offinterlineskip\halign{\hfil$\scriptscriptstyle##$\hfil\cr
<\cr\sim\cr}}}}}

\def\ga{\mathrel{\mathchoice {\vcenter{\offinterlineskip\halign{\hfil
$\displaystyle##$\hfil\cr>\cr\sim\cr}}}
{\vcenter{\offinterlineskip\halign{\hfil$\textstyle##$\hfil\cr
>\cr\sim\cr}}}
{\vcenter{\offinterlineskip\halign{\hfil$\scriptstyle##$\hfil\cr
>\cr\sim\cr}}}
{\vcenter{\offinterlineskip\halign{\hfil$\scriptscriptstyle##$\hfil\cr
>\cr\sim\cr}}}}}

\def\sun{\hbox{$\odot$}}

\def\diameter{{\ifmmode\mathchoice
{\ooalign{\hfil\hbox{$\displaystyle/$}\hfil\crcr
{\hbox{$\displaystyle\mathchar"20D$}}}}
{\ooalign{\hfil\hbox{$\textstyle/$}\hfil\crcr
{\hbox{$\textstyle\mathchar"20D$}}}}
{\ooalign{\hfil\hbox{$\scriptstyle/$}\hfil\crcr
{\hbox{$\scriptstyle\mathchar"20D$}}}}
{\ooalign{\hfil\hbox{$\scriptscriptstyle/$}\hfil\crcr
{\hbox{$\scriptscriptstyle\mathchar"20D$}}}}
\else{\ooalign{\hfil/\hfil\crcr\mathhexbox20D}}%
\fi}}

\def\sq{\ifmmode\squareforqed\else{\unskip\nobreak\hfil
\penalty50\hskip1em\null\nobreak\hfil\squareforqed
\parfillskip=0pt\finalhyphendemerits=0\endgraf}\fi}
\def\squareforqed{\hbox{\rlap{$\sqcap$}$\sqcup$}}


\def\bbbc{{\mathchoice {\setbox0=\hbox{$\displaystyle\rm C$}\hbox{\hbox
to0pt{\kern0.4\wd0\vrule height0.9\ht0\hss}\box0}}
{\setbox0=\hbox{$\textstyle\rm C$}\hbox{\hbox
to0pt{\kern0.4\wd0\vrule height0.9\ht0\hss}\box0}}
{\setbox0=\hbox{$\scriptstyle\rm C$}\hbox{\hbox
to0pt{\kern0.4\wd0\vrule height0.9\ht0\hss}\box0}}
{\setbox0=\hbox{$\scriptscriptstyle\rm C$}\hbox{\hbox
to0pt{\kern0.4\wd0\vrule height0.9\ht0\hss}\box0}}}}
\def\bbbq{{\mathchoice {\setbox0=\hbox{$\displaystyle\rm
Q$}\hbox{\raise
0.15\ht0\hbox to0pt{\kern0.4\wd0\vrule height0.8\ht0\hss}\box0}}
{\setbox0=\hbox{$\textstyle\rm Q$}\hbox{\raise
0.15\ht0\hbox to0pt{\kern0.4\wd0\vrule height0.8\ht0\hss}\box0}}
{\setbox0=\hbox{$\scriptstyle\rm Q$}\hbox{\raise
0.15\ht0\hbox to0pt{\kern0.4\wd0\vrule height0.7\ht0\hss}\box0}}
{\setbox0=\hbox{$\scriptscriptstyle\rm Q$}\hbox{\raise
0.15\ht0\hbox to0pt{\kern0.4\wd0\vrule height0.7\ht0\hss}\box0}}}}
\def\bbbt{{\mathchoice {\setbox0=\hbox{$\displaystyle\rm
T$}\hbox{\hbox to0pt{\kern0.3\wd0\vrule height0.9\ht0\hss}\box0}}
{\setbox0=\hbox{$\textstyle\rm T$}\hbox{\hbox
to0pt{\kern0.3\wd0\vrule height0.9\ht0\hss}\box0}}
{\setbox0=\hbox{$\scriptstyle\rm T$}\hbox{\hbox
to0pt{\kern0.3\wd0\vrule height0.9\ht0\hss}\box0}}
{\setbox0=\hbox{$\scriptscriptstyle\rm T$}\hbox{\hbox
to0pt{\kern0.3\wd0\vrule height0.9\ht0\hss}\box0}}}}
\def\bbbs{{\mathchoice
{\setbox0=\hbox{$\displaystyle     \rm S$}\hbox{\raise0.5\ht0\hbox
to0pt{\kern0.35\wd0\vrule height0.45\ht0\hss}\hbox
to0pt{\kern0.55\wd0\vrule height0.5\ht0\hss}\box0}}
{\setbox0=\hbox{$\textstyle        \rm S$}\hbox{\raise0.5\ht0\hbox
to0pt{\kern0.35\wd0\vrule height0.45\ht0\hss}\hbox
to0pt{\kern0.55\wd0\vrule height0.5\ht0\hss}\box0}}
{\setbox0=\hbox{$\scriptstyle      \rm S$}\hbox{\raise0.5\ht0\hbox
to0pt{\kern0.35\wd0\vrule height0.45\ht0\hss}\raise0.05\ht0\hbox
to0pt{\kern0.5\wd0\vrule height0.45\ht0\hss}\box0}}
{\setbox0=\hbox{$\scriptscriptstyle\rm S$}\hbox{\raise0.5\ht0\hbox
to0pt{\kern0.4\wd0\vrule height0.45\ht0\hss}\raise0.05\ht0\hbox
to0pt{\kern0.55\wd0\vrule height0.45\ht0\hss}\box0}}}}
\def\bbbz{{\mathchoice {\hbox{$\sf\textstyle Z\kern-0.4em Z$}}
{\hbox{$\sf\textstyle Z\kern-0.4em Z$}}
{\hbox{$\sf\scriptstyle Z\kern-0.3em Z$}}
{\hbox{$\sf\scriptscriptstyle Z\kern-0.2em Z$}}}}


\def\Nulle{0} 
\def\Afe{1}   
\def\Hae{2}   
\def\Hbe{3}   
\def\Hce{4}   
\def\Hde{5}   


\newcount\LastMac       \LastMac=\Nulle

\newskip\half      \half=5.5pt plus 1.5pt minus 2.25pt
\newskip\one       \one=11pt plus 3pt minus 5.5pt
\newskip\onehalf   \onehalf=16.5pt plus 5.5pt minus 8.25pt
\newskip\two       \two=22pt plus 5.5pt minus 11pt

\def\Half{\addvspace{\half}}
\def\One{\addvspace{\one}}
\def\OneHalf{\addvspace{\onehalf}}
\def\Two{\addvspace{\two}}

\def\Raggedright{
  \rightskip=\z@ plus \hsize\relax
}

\def\Fullout{
  \rightskip=\z@\relax
}

\def\Hang#1#2{
  \hangindent=#1%
  \hangafter=#2\relax
}


\newif\ifsp@page
\def\pagestyle#1{\csname ps@#1\endcsname}
\def\thispagestyle#1{\global\sp@pagetrue\gdef\sp@type{#1}}

\def\ps@titlepage{%
  \def\@oddhead{\eightpoint\noindent \the\CatchLine
    \ifprod@font\else\qquad Printed\ \today\qquad
      (MN plain \TeX\ macros\ v\@version)\fi \hfil}%
  \let\@evenhead=\@oddhead
  \def\@oddfoot{\eightpoint\copyright\ \@pubyear\ RAS\hfil}%
  \def\@evenfoot{\hfil\eightpoint\noindent\copyright\ \@pubyear\ RAS}%
}

\def\ps@headings{%
  \def\@oddhead{\elevenpoint\it\noindent
    \hfill\the\RightHeader\hskip1.5em\rm\folio}%
  \def\@evenhead{\elevenpoint\noindent
    \folio\hskip1.5em\it\the\LeftHeader\hfill}%
  \def\@oddfoot{\eightpoint\noindent\copyright\ \@pubyear\ RAS,
    MNRAS {\bf \@volume}, \@pagerange\hfil}%
  \def\@evenfoot{\hfil\eightpoint\copyright\ \@pubyear\ RAS,
    MNRAS {\bf \@volume}, \@pagerange}%
}

\def\ps@plate{%
  \def\@oddhead{\eightpoint\noindent\plt@cap\hfil}%
  \def\@evenhead{\eightpoint\noindent\plt@cap\hfil}%
  \def\@oddfoot{\eightpoint\noindent\copyright\ \@pubyear\ RAS,
    MNRAS {\bf \@volume}, \@pagerange\hfil}%
  \def\@evenfoot{\hfil\eightpoint\copyright\ \@pubyear\ RAS,
    MNRAS {\bf \@volume}, \@pagerange}%
}



\def\title#1{
  \bgroup
    \vbox to 8pt{\vss}%
    \seventeenpoint
    \Raggedright
    \noindent \strut{\bf #1}\par
  \egroup
}

\def\author#1{
  \bgroup
    \ifnum\LastMac=\Afe \OneHalf\else \vskip 21pt\fi
    \fourteenpoint
    \Raggedright
    \noindent \strut #1\par
    \vskip 3pt%
  \egroup
}

\def\affiliation#1{
  \bgroup
    \vskip -4pt%
    \eightpoint
    \Raggedright
    \noindent \strut {\it #1}\par
  \egroup
  \LastMac=\Afe\relax
}

\def\acceptedline#1{
  \bgroup
    \Two
    \eightpoint
    \Raggedright
    \noindent \strut #1\par
  \egroup
}

\long\def\abstract#1{%
  \bgroup
    \vskip 20pt%
    \leftskip 11pc\rightskip\z@
    \noindent{\ninebf ABSTRACT}\par
    \tenpoint
    \Fullout
    \noindent #1\par
  \egroup
}

\long\def\keywords#1{
  \bgroup
    \Half
    \leftskip 11pc\rightskip\z@
    \tenpoint
    \Fullout
    \noindent\hbox{\bf Key words:}\ #1\par
  \egroup
}


\def\maketitle{%
  \EndOpening
  \ifsinglecol \else \MakePage\fi
}


\def\pageoffset#1#2{\hoffset=#1\relax\voffset=#2\relax}


\def\@nameuse#1{\csname #1\endcsname}
\def\arabic#1{\@arabic{\@nameuse{#1}}}
\def\alph#1{\@alph{\@nameuse{#1}}}
\def\Alph#1{\@Alph{\@nameuse{#1}}}
\def\@arabic#1{\number #1}
\def\@Alph#1{\ifcase#1\or A\or B\or C\or D\else\@Ialph{#1}\fi}
\def\@Ialph#1{\ifcase#1\or \or \or \or \or E\or F\or G\or H\or I\or J\or
   K\or L\or M\or N\or O\or P\or Q\or R\or S\or T\or U\or V\or W\or X\or
   Y\or Z\else\errmessage{Counter out of range}\fi}
\def\@alph#1{\ifcase#1\or a\or b\or c\or d\else\@ialph{#1}\fi}
\def\@ialph#1{\ifcase#1\or \or \or \or \or e\or f\or g\or h\or i\or j\or
   k\or l\or m\or n\or o\or p\or q\or r\or s\or t\or u\or v\or w\or x\or y\or
   z\else\errmessage{Counter out of range}\fi}


\newcount\Eqnno
\newcount\SubEqnno

\def\theeq{\arabic{Eqnno}}
\def\thesubeq{\alph{SubEqnno}}

\def\stepeq{\relax
  \global\SubEqnno \z@
  \global\advance\Eqnno \@ne\relax
  {\rm (\theeq)}%
}

\def\startsubeq{\relax
  \global\SubEqnno \z@
  \global\advance\Eqnno \@ne\relax
  \stepsubeq
}

\def\stepsubeq{\relax
  \global\advance\SubEqnno \@ne\relax
  {\rm (\theeq\thesubeq)}%
}


\newcount\Sec        
\newcount\SecSec
\newcount\SecSecSec

\def\thesection{\arabic{Sec}}
\def\thesubsection{\thesection.\arabic{SecSec}}
\def\thesubsubsection{\thesubsection.\arabic{SecSecSec}}

\Sec=\z@

\def\:{\let\@sptoken= } \:  
\def\:{\@xifnch} \expandafter\def\: {\futurelet\@tempc\@ifnch}

\def\@ifnextchar#1#2#3{%
  \let\@tempMACe #1%
  \def\@tempMACa{#2}%
  \def\@tempMACb{#3}%
  \futurelet \@tempMACc\@ifnch%
}

\def\@ifnch{%
\ifx \@tempMACc \@sptoken%
  \let\@tempMACd\@xifnch%
\else%
  \ifx \@tempMACc \@tempMACe%
    \let\@tempMACd\@tempMACa%
  \else%
    \let\@tempMACd\@tempMACb%
  \fi%
\fi%
\@tempMACd%
}

\def\@ifstar#1#2{\@ifnextchar *{\def\@tempMACa*{#1}\@tempMACa}{#2}}

\newskip\@tempskipb

\def\addvspace#1{%
  \ifvmode\else \endgraf\fi%
  \ifdim\lastskip=\z@%
    \vskip #1\relax%
  \else%
    \@tempskipb#1\relax\@xaddvskip%
  \fi%
}

\def\@xaddvskip{%
  \ifdim\lastskip<\@tempskipb%
    \vskip-\lastskip%
    \vskip\@tempskipb\relax%
  \else%
    \ifdim\@tempskipb<\z@%
      \ifdim\lastskip<\z@ \else%
        \advance\@tempskipb\lastskip%
        \vskip-\lastskip\vskip\@tempskipb%
      \fi%
    \fi%
  \fi%
}

\newskip\@tmpSKIP

\def\addpen#1{%
  \ifvmode
    \if@nobreak
    \else
      \ifdim\lastskip=\z@
        \penalty#1\relax
      \else
        \@tmpSKIP=\lastskip
        \vskip -\lastskip
        \penalty#1\vskip\@tmpSKIP
      \fi
    \fi
  \fi
}

\newcount\@clubpen   \@clubpen=\clubpenalty
\newif\if@nobreak    \@nobreakfalse

\def\@noafterindent{%
  \global\@nobreaktrue
  \everypar{\if@nobreak
              \global\@nobreakfalse
              \clubpenalty \@M
              {\setbox\z@\lastbox}%
              \LastMac=\Nulle\relax%
            \else
              \clubpenalty \@clubpen
              \everypar{}%
            \fi}%
}

\newcount\gds@cbrk   \gds@cbrk=-300

\def\@nohdbrk{\interlinepenalty \@M\relax}

\let\@par=\par
\def\@restorepar{\def\par{\@par}}

\newif\if@endpe   \@endpefalse
 
\def\@doendpe{\@endpetrue \@nobreakfalse \LastMac=\Nulle\relax%
     \def\par{\@restorepar\everypar{}\par\@endpefalse}%
              \everypar{\setbox\z@\lastbox\everypar{}\@endpefalse}%
}

\def\section{\@ifstar{\@ssection}{\@section}}

\def\@section#1{
  \if@nobreak
    \everypar{}%
    \ifnum\LastMac=\Hae \addvspace{\half}\fi
  \else
    \addpen{\gds@cbrk}%
    \addvspace{\two}%
  \fi
  \bgroup
    \ninepoint\bf
    \Raggedright
    \global\advance\Sec \@ne
    \ifappendix
      \global\Eqnno=\z@ \global\SubEqnno=\z@\relax
      \def\ch@ck{#1}%
      \ifx\ch@ck\empty \def\c@lon{}\else\def\c@lon{:}\fi
      \noindent\@nohdbrk APPENDIX\ \thesection\c@lon\hskip 0.5em%
        \uppercase{#1}\par
    \else
      \noindent\@nohdbrk\thesection\hskip 1pc \uppercase{#1}\par
    \fi
    \global\SecSec=\z@
  \egroup
  \nobreak
  \vskip\half
  \nobreak
  \@noafterindent
  \LastMac=\Hae\relax
}

\def\@ssection#1{
  \if@nobreak
    \everypar{}%
    \ifnum\LastMac=\Hae \addvspace{\half}\fi
  \else
    \addpen{\gds@cbrk}%
    \addvspace{\two}%
  \fi
  \bgroup
    \ninepoint\bf
    \Raggedright
    \noindent\@nohdbrk\uppercase{#1}\par
  \egroup
  \nobreak
  \vskip\half
  \nobreak
  \@noafterindent
  \LastMac=\Hae\relax
}

\def\subsection{\@ifstar{\@ssubsection}{\@subsection}}

\def\@subsection#1{
  \if@nobreak
    \everypar{}%
    \ifnum\LastMac=\Hae \addvspace{1pt plus 1pt minus .5pt}\fi
  \else
    \addpen{\gds@cbrk}%
    \addvspace{\onehalf}%
  \fi
  \bgroup
    \ninepoint\bf
    \Raggedright
    \global\advance\SecSec \@ne
    \noindent\@nohdbrk\thesubsection \hskip 1pc\relax #1\par
    \global\SecSecSec=\z@
  \egroup
  \nobreak
  \vskip\half
  \nobreak
  \@noafterindent
  \LastMac=\Hbe\relax
}

\def\@ssubsection#1{
  \if@nobreak
    \everypar{}%
    \ifnum\LastMac=\Hae \addvspace{1pt plus 1pt minus .5pt}\fi
  \else
    \addpen{\gds@cbrk}%
    \addvspace{\onehalf}%
  \fi
  \bgroup
    \ninepoint\bf
    \Raggedright
    \noindent\@nohdbrk #1\par
  \egroup
  \nobreak
  \vskip\half
  \nobreak
  \@noafterindent
  \LastMac=\Hbe\relax
}

\def\subsubsection{\@ifstar{\@ssubsubsection}{\@subsubsection}}

\def\@subsubsection#1{
  \if@nobreak
    \everypar{}%
    \ifnum\LastMac=\Hbe \addvspace{1pt plus 1pt minus .5pt}\fi
  \else
    \addpen{\gds@cbrk}%
    \addvspace{\onehalf}%
  \fi
  \bgroup
    \ninepoint\it
    \Raggedright
    \global\advance\SecSecSec \@ne
    \noindent\@nohdbrk\thesubsubsection \hskip 1pc\relax #1\par
  \egroup
  \nobreak
  \vskip\half
  \nobreak
  \@noafterindent
  \LastMac=\Hce\relax
}

\def\@ssubsubsection#1{
  \if@nobreak
    \everypar{}%
    \ifnum\LastMac=\Hbe \addvspace{1pt plus 1pt minus .5pt}\fi
  \else
    \addpen{\gds@cbrk}%
    \addvspace{\onehalf}%
  \fi
  \bgroup
    \ninepoint\it
    \Raggedright
    \noindent\@nohdbrk #1\par
  \egroup
  \nobreak
  \vskip\half
  \nobreak
  \@noafterindent
  \LastMac=\Hce\relax
}

\def\paragraph#1{
  \if@nobreak
    \everypar{}%
  \else
    \addpen{\gds@cbrk}%
    \addvspace{\one}%
  \fi%
  \bgroup%
    \ninepoint\it
    \noindent #1\ \nobreak%
  \egroup
  \LastMac=\Hde\relax
  \ignorespaces
}


\newif\ifappendix

\def\appendix{%
  \global\appendixtrue
  \def\thesection{\Alph{Sec}}%
  \def\thesubsection{\thesection\arabic{SecSec}}%
  \def\theeq{\thesection\arabic{Eqnno}}%
  \Sec=\z@ \SecSec=\z@ \SecSecSec=\z@ \Eqnno=\z@ \SubEqnno=\z@\relax
}




\def\beginlist{%
  \par\if@nobreak \else\addvspace{\half}\fi%
  \bgroup%
    \ninepoint
    \let\item=\list@item%
}

\def\list@item{%
  \par\noindent\hskip 1em\relax%
  \ignorespaces%
}

\def\endlist{\par\egroup\addvspace{\half}\@doendpe}


\def\beginrefs{%
  \par
  \bgroup
    \eightpoint
    \Fullout
    \let\bibitem=\bib@item
}

\def\bib@item{%
  \par\parindent=1.5em\Hang{1.5em}{1}%
  \everypar={\Hang{1.5em}{1}\ignorespaces}%
  \noindent\ignorespaces
}

\def\endrefs{\par\egroup\@doendpe}


\newtoks\CatchLine

\def\@journal{Mon.\ Not.\ R.\ Astron.\ Soc.\ }  
\def\@pubyear{1994}        
\def\@pagerange{000--000}  
\def\@volume{000}          
\def\@microfiche{}         %

\def\pubyear#1{\gdef\@pubyear{#1}\@makecatchline}
\def\pagerange#1{\gdef\@pagerange{#1}\@makecatchline}
\def\volume#1{\gdef\@volume{#1}\@makecatchline}
\def\microfiche#1{\gdef\@microfiche{and Microfiche\ #1}\@makecatchline}

\def\@makecatchline{%
  \global\CatchLine{%
    {\rm \@journal {\bf \@volume},\ \@pagerange\ (\@pubyear)\ \@microfiche}}%
}

\@makecatchline 

\newtoks\LeftHeader
\def\shortauthor#1{
  \global\LeftHeader{#1}%
}

\newtoks\RightHeader
\def\shorttitle#1{
  \global\RightHeader{#1}%
}

\def\PageHead{
  \begingroup
    \ifsp@page
      \csname ps@\sp@type\endcsname
    \fi
    \ifodd\pageno
      \let\the@head=\@oddhead
    \else
      \let\the@head=\@evenhead
    \fi
    \vbox to \z@{\vskip-22.5\p@%
      \hbox to \PageWidth{\vbox to8.5\p@{}%
        \the@head
      }%
    \vss}%
  \endgroup
  \nointerlineskip
}

\gdef\PageFoot{%
  \nointerlineskip%
  \begingroup
  \ifsp@page
    \csname ps@\sp@type\endcsname
    \global\sp@pagefalse
  \fi
  \vbox to 22pt{\vfil%
    \hbox to \PageWidth{%
      \eightpoint\strut\noindent
      \ifodd\pageno
        \@oddfoot
      \else
        \@evenfoot
      \fi
    }%
  }%
  \endgroup
}

\def\today{%
  \number\day\space
  \ifcase\month\or January\or February\or March\or April\or May\or June\or
    July\or August\or September\or October\or November\or December\fi
  \space\number\year%
}

\def\authorcomment#1{%
  \gdef\PageFoot{%
    \nointerlineskip%
    \vbox to 20pt{\vfil%
      \hbox to \PageWidth{\elevenpoint\noindent \hfil #1 \hfil}}%
  }%
}


\newif\ifplate@page
\newbox\plt@box

\def\beginplatepage{%
  \let\plate=\plate@head
  \let\caption=\fig@caption
  \global\setbox\plt@box=\vbox\bgroup
  \TEMPDIMEN=\PageWidth 
  \hsize=\PageWidth\relax
}

\def\endplatepage{\par\egroup\global\plate@pagetrue}
\def\plate@head#1{\gdef\plt@cap{#1}}


\def\letters{%
  \gdef\folio{\ifnum\pageno<\z@ L\romannumeral-\pageno
    \else L\number\pageno \fi}%
}


\newdimen\mathindent

\global\mathindent=\z@
\global\everydisplay{\global\@dspwd=\displaywidth\displaysetup}


\def\@displaylines#1{
  {}$\displ@y\hbox{\vbox{\halign{$\@lign\hfil\displaystyle##\hfil$\crcr
  #1\crcr}}}${}%
}

\def\@eqalign#1{\null\vcenter{\openup\jot\m@th
  \ialign{\strut\hfil$\displaystyle{##}$&$\displaystyle{{}##}$\hfil
      \crcr#1\crcr}}%
}

\def\@eqalignno#1{
  \global\advance\@dspwd by -\mathindent%
  {}$\displ@y\hbox{\vbox{\halign to\@dspwd%
  {\hfil$\@lign\displaystyle{##}$\tabskip\z@skip
  &$\@lign\displaystyle{{}##}$\hfil\tabskip\centering
  &\llap{$\@lign##$}\tabskip\z@skip\crcr
  #1\crcr}}}${}%
}


\global\let\displaylines=\@displaylines
\global\let\eqalign=\@eqalign
\global\let\eqalignno=\@eqalignno
\global\let\leqalignno=\@eqalignno

\newdimen\@dspwd   \@dspwd=\z@
\newif\if@eqno
\newif\if@leqno
\newtoks\@eqn
\newtoks\@eq

\def\displaysetup#1$${\displaytest#1\eqno\eqno\displaytest}

\def\displaytest#1\eqno#2\eqno#3\displaytest{%
 \if!#3!\ldisplaytest#1\leqno\leqno\ldisplaytest
 \else\@eqnotrue\@leqnofalse\@eqn={#2}\@eq={#1}\fi
 \generaldisplay$$}

\def\ldisplaytest#1\leqno#2\leqno#3\ldisplaytest{%
\@eq={#1}%
 \if!#3!\@eqnofalse\else\@eqnotrue\@leqnotrue
  \@eqn={#2}\fi}

\def\generaldisplay{%
  \if@eqno
    \if@leqno
      \hbox to \displaywidth{\noindent
        \rlap{$\displaystyle\the\@eqn$}%
        \hskip\mathindent$\displaystyle\the\@eq$\hfil}%
    \else
      \hbox to \displaywidth{\noindent
        \hskip\mathindent
        $\displaystyle\the\@eq$\hfil$\displaystyle\the\@eqn$}%
    \fi
  \else
    \hbox to \displaywidth{\noindent
      \hskip\mathindent$\displaystyle\the\@eq$\hfil}%
  \fi
}


\def\@notice{%
  \par\Two%
  \noindent{\b@ls{11pt}\ninerm This paper has been produced using the
    Royal Astronomical Society/Blackwell Science \TeX\ macros.\par}%
}

\outer\def\bye{\@notice\par\vfill\supereject\end}


\def\start@mess{%
  Monthly notices of the RAS journal style (\@typeface)\space
    v\@version,\space \@verdate.%
}

\everyjob{\Warn{\start@mess}}



\newif\if@debug \@debugfalse  

\def\Print#1{\if@debug\immediate\write16{#1}\else \fi}
\def\Warn#1{\immediate\write16{#1}}
\def\wlog#1{}

\newcount\Iteration 

\def\Single{0} \def\Double{1}                 
\def\Figure{0} \def\Table{1}                  

\def\InStack{0}  
\def\InZoneA{1}
\def\InZoneB{2}
\def\InZoneC{3}

\newcount\TEMPCOUNT 
\newdimen\TEMPDIMEN 
\newbox\TEMPBOX     
\newbox\VOIDBOX     

\newcount\LengthOfStack 
\newcount\MaxItems      
\newcount\StackPointer
\newcount\Point         
\newcount\NextFigure    
\newcount\NextTable     
\newcount\NextItem      

\newcount\StatusStack   
\newcount\NumStack      
\newcount\TypeStack     
\newcount\SpanStack     
\newcount\BoxStack      

\newcount\ItemSTATUS    
\newcount\ItemNUMBER    
\newcount\ItemTYPE      
\newcount\ItemSPAN      
\newbox\ItemBOX         
\newdimen\ItemSIZE      

\newdimen\PageHeight    
\newdimen\TextLeading   
\newdimen\Feathering    
\newcount\LinesPerPage  
\newdimen\ColumnWidth   
\newdimen\ColumnGap     
\newdimen\PageWidth     
\newdimen\BodgeHeight   
\newcount\Leading       

\newdimen\ZoneBSize  
\newdimen\TextSize   
\newbox\ZoneABOX     
\newbox\ZoneBBOX     
\newbox\ZoneCBOX     

\newif\ifFirstSingleItem
\newif\ifFirstZoneA
\newif\ifMakePageInComplete
\newif\ifMoreFigures \MoreFiguresfalse 
\newif\ifMoreTables  \MoreTablesfalse  

\newif\ifFigInZoneB 
\newif\ifFigInZoneC 
\newif\ifTabInZoneB 
\newif\ifTabInZoneC

\newif\ifZoneAFullPage

\newbox\MidBOX    
\newbox\LeftBOX
\newbox\RightBOX
\newbox\PageBOX   

\newif\ifLeftCOL  
\LeftCOLtrue

\newdimen\ZoneBAdjust

\newcount\ItemFits
\def\Yes{1}
\def\No{2}


\MaxItems=15
\NextFigure=\z@        
\NextTable=\@ne

\BodgeHeight=6pt
\TextLeading=11pt    
\Leading=11
\Feathering=\z@      
\LinesPerPage=61     
\topskip=\TextLeading
\ColumnWidth=20pc    
\ColumnGap=2pc       

\newskip\ItemSepamount  
\ItemSepamount=\TextLeading plus \TextLeading minus 4pt

\parskip=\z@ plus .1pt
\parindent=18pt
\widowpenalty=\z@
\clubpenalty=10000
\tolerance=1500
\hbadness=1500
\abovedisplayskip=6pt plus 2pt minus 1pt
\belowdisplayskip=6pt plus 2pt minus 1pt
\abovedisplayshortskip=6pt plus 2pt minus 1pt
\belowdisplayshortskip=6pt plus 2pt minus 1pt

\frenchspacing

\ninepoint 

\PageHeight=682pt
\PageWidth=2\ColumnWidth
\advance\PageWidth by \ColumnGap

\pagestyle{headings}




\newcount\DUMMY \StatusStack=\allocationnumber
\newcount\DUMMY \newcount\DUMMY \newcount\DUMMY 
\newcount\DUMMY \newcount\DUMMY \newcount\DUMMY 
\newcount\DUMMY \newcount\DUMMY \newcount\DUMMY
\newcount\DUMMY \newcount\DUMMY \newcount\DUMMY 
\newcount\DUMMY \newcount\DUMMY \newcount\DUMMY

\newcount\DUMMY \NumStack=\allocationnumber
\newcount\DUMMY \newcount\DUMMY \newcount\DUMMY 
\newcount\DUMMY \newcount\DUMMY \newcount\DUMMY 
\newcount\DUMMY \newcount\DUMMY \newcount\DUMMY 
\newcount\DUMMY \newcount\DUMMY \newcount\DUMMY 
\newcount\DUMMY \newcount\DUMMY \newcount\DUMMY

\newcount\DUMMY \TypeStack=\allocationnumber
\newcount\DUMMY \newcount\DUMMY \newcount\DUMMY 
\newcount\DUMMY \newcount\DUMMY \newcount\DUMMY 
\newcount\DUMMY \newcount\DUMMY \newcount\DUMMY 
\newcount\DUMMY \newcount\DUMMY \newcount\DUMMY 
\newcount\DUMMY \newcount\DUMMY \newcount\DUMMY

\newcount\DUMMY \SpanStack=\allocationnumber
\newcount\DUMMY \newcount\DUMMY \newcount\DUMMY 
\newcount\DUMMY \newcount\DUMMY \newcount\DUMMY 
\newcount\DUMMY \newcount\DUMMY \newcount\DUMMY 
\newcount\DUMMY \newcount\DUMMY \newcount\DUMMY 
\newcount\DUMMY \newcount\DUMMY \newcount\DUMMY

\newbox\DUMMY   \BoxStack=\allocationnumber
\newbox\DUMMY   \newbox\DUMMY \newbox\DUMMY 
\newbox\DUMMY   \newbox\DUMMY \newbox\DUMMY 
\newbox\DUMMY   \newbox\DUMMY \newbox\DUMMY 
\newbox\DUMMY   \newbox\DUMMY \newbox\DUMMY 
\newbox\DUMMY   \newbox\DUMMY \newbox\DUMMY

\def\wlog{\immediate\write\m@ne}


\def\GetItemAll#1{%
 \GetItemSTATUS{#1}
 \GetItemNUMBER{#1}
 \GetItemTYPE{#1}
 \GetItemSPAN{#1}
 \GetItemBOX{#1}
}

\def\GetItemSTATUS#1{%
 \Point=\StatusStack
 \advance\Point by #1
 \global\ItemSTATUS=\count\Point
}

\def\GetItemNUMBER#1{%
 \Point=\NumStack
 \advance\Point by #1
 \global\ItemNUMBER=\count\Point
}

\def\GetItemTYPE#1{%
 \Point=\TypeStack
 \advance\Point by #1
 \global\ItemTYPE=\count\Point
}

\def\GetItemSPAN#1{%
 \Point\SpanStack
 \advance\Point by #1
 \global\ItemSPAN=\count\Point
}

\def\GetItemBOX#1{%
 \Point=\BoxStack
 \advance\Point by #1
 \global\setbox\ItemBOX=\vbox{\copy\Point}
 \global\ItemSIZE=\ht\ItemBOX
 \global\advance\ItemSIZE by \dp\ItemBOX
 \TEMPCOUNT=\ItemSIZE
 \divide\TEMPCOUNT by \Leading
 \divide\TEMPCOUNT by 65536
 \advance\TEMPCOUNT \@ne
 \ItemSIZE=\TEMPCOUNT pt
 \global\multiply\ItemSIZE by \Leading
}


\def\JoinStack{%
 \ifnum\LengthOfStack=\MaxItems 
  \Warn{WARNING: Stack is full...some items will be lost!}
 \else
  \Point=\StatusStack
  \advance\Point by \LengthOfStack
  \global\count\Point=\ItemSTATUS
  \Point=\NumStack
  \advance\Point by \LengthOfStack
  \global\count\Point=\ItemNUMBER
  \Point=\TypeStack
  \advance\Point by \LengthOfStack
  \global\count\Point=\ItemTYPE
  \Point\SpanStack
  \advance\Point by \LengthOfStack
  \global\count\Point=\ItemSPAN
  \Point=\BoxStack
  \advance\Point by \LengthOfStack
  \global\setbox\Point=\vbox{\copy\ItemBOX}
  \global\advance\LengthOfStack \@ne
  \ifnum\ItemTYPE=\Figure 
   \global\MoreFigurestrue
  \else
   \global\MoreTablestrue
  \fi
 \fi
}


\def\LeaveStack#1{%
 {\Iteration=#1
 \loop
 \ifnum\Iteration<\LengthOfStack
  \advance\Iteration \@ne
  \GetItemSTATUS{\Iteration}
   \advance\Point by \m@ne
   \global\count\Point=\ItemSTATUS
  \GetItemNUMBER{\Iteration}
   \advance\Point by \m@ne
   \global\count\Point=\ItemNUMBER
  \GetItemTYPE{\Iteration}
   \advance\Point by \m@ne
   \global\count\Point=\ItemTYPE
  \GetItemSPAN{\Iteration}
   \advance\Point by \m@ne
   \global\count\Point=\ItemSPAN
  \GetItemBOX{\Iteration}
   \advance\Point by \m@ne
   \global\setbox\Point=\vbox{\copy\ItemBOX}
 \repeat}
 \global\advance\LengthOfStack by \m@ne
}


\newif\ifStackNotClean

\def\CleanStack{%
 \StackNotCleantrue
 {\Iteration=\z@
  \loop
   \ifStackNotClean
    \GetItemSTATUS{\Iteration}
    \ifnum\ItemSTATUS=\InStack
     \advance\Iteration \@ne
     \else
      \LeaveStack{\Iteration}
    \fi
   \ifnum\LengthOfStack<\Iteration
    \StackNotCleanfalse
   \fi
 \repeat}
}


\def\FindItem#1#2{%
 \global\StackPointer=\m@ne 
 {\Iteration=\z@
  \loop
  \ifnum\Iteration<\LengthOfStack
   \GetItemSTATUS{\Iteration}
   \ifnum\ItemSTATUS=\InStack
    \GetItemTYPE{\Iteration}
    \ifnum\ItemTYPE=#1
     \GetItemNUMBER{\Iteration}
     \ifnum\ItemNUMBER=#2
      \global\StackPointer=\Iteration
      \Iteration=\LengthOfStack 
     \fi
    \fi
   \fi
  \advance\Iteration \@ne
 \repeat}
}


\def\FindNext{%
 \global\StackPointer=\m@ne 
 {\Iteration=\z@
  \loop
  \ifnum\Iteration<\LengthOfStack
   \GetItemSTATUS{\Iteration}
   \ifnum\ItemSTATUS=\InStack
    \GetItemTYPE{\Iteration}
   \ifnum\ItemTYPE=\Figure
    \ifMoreFigures
      \global\NextItem=\Figure
      \global\StackPointer=\Iteration
      \Iteration=\LengthOfStack 
    \fi
   \fi
   \ifnum\ItemTYPE=\Table
    \ifMoreTables
      \global\NextItem=\Table
      \global\StackPointer=\Iteration
      \Iteration=\LengthOfStack 
    \fi
   \fi
  \fi
  \advance\Iteration \@ne
 \repeat}
}


\def\ChangeStatus#1#2{%
 \Point=\StatusStack
 \advance\Point by #1
 \global\count\Point=#2
}



\def\Zone{\InZoneA}

\ZoneBAdjust=\z@

\def\MakePage{
 \global\ZoneBSize=\PageHeight
 \global\TextSize=\ZoneBSize
 \global\ZoneAFullPagefalse
 \global\topskip=\TextLeading
 \MakePageInCompletetrue
 \MoreFigurestrue
 \MoreTablestrue
 \FigInZoneBfalse
 \FigInZoneCfalse
 \TabInZoneBfalse
 \TabInZoneCfalse
 \global\FirstSingleItemtrue
 \global\FirstZoneAtrue
 \global\setbox\ZoneABOX=\box\VOIDBOX
 \global\setbox\ZoneBBOX=\box\VOIDBOX
 \global\setbox\ZoneCBOX=\box\VOIDBOX
 \loop
  \ifMakePageInComplete
 \FindNext
 \ifnum\StackPointer=\m@ne
  \NextItem=\m@ne
  \MoreFiguresfalse
  \MoreTablesfalse
 \fi
 \ifnum\NextItem=\Figure
   \FindItem{\Figure}{\NextFigure}
   \ifnum\StackPointer=\m@ne \global\MoreFiguresfalse
   \else
    \GetItemSPAN{\StackPointer}
    \ifnum\ItemSPAN=\Single \def\Zone{\InZoneB}\relax
     \ifFigInZoneC \global\MoreFiguresfalse\fi
    \else
     \def\Zone{\InZoneA}
     \ifFigInZoneB \def\Zone{\InZoneC}\fi
    \fi
   \fi
   \ifMoreFigures\Print{}\FigureItems\fi
 \fi
\ifnum\NextItem=\Table
   \FindItem{\Table}{\NextTable}
   \ifnum\StackPointer=\m@ne \global\MoreTablesfalse
   \else
    \GetItemSPAN{\StackPointer}
    \ifnum\ItemSPAN=\Single\relax
     \ifTabInZoneC \global\MoreTablesfalse\fi
    \else
     \def\Zone{\InZoneA}
     \ifTabInZoneB \def\Zone{\InZoneC}\fi
    \fi
   \fi
   \ifMoreTables\Print{}\TableItems\fi
 \fi
   \MakePageInCompletefalse 
   \ifMoreFigures\MakePageInCompletetrue\fi
   \ifMoreTables\MakePageInCompletetrue\fi
 \repeat
 \ifZoneAFullPage
  \global\TextSize=\z@
  \global\ZoneBSize=\z@
  \global\vsize=\z@\relax
  \global\topskip=\z@\relax
  \vbox to \z@{\vss}
  \eject
 \else
 \global\advance\ZoneBSize by -\ZoneBAdjust
 \global\vsize=\ZoneBSize
 \global\hsize=\ColumnWidth
 \global\ZoneBAdjust=\z@
 \ifdim\TextSize<23pt
 \Warn{}
 \Warn{* Making column fall short: TextSize=\the\TextSize *}
 \vskip-\lastskip\eject\fi
 \fi
}

\def\MakeRightCol{
 \global\TextSize=\ZoneBSize
 \MakePageInCompletetrue
 \MoreFigurestrue
 \MoreTablestrue
 \global\FirstSingleItemtrue
 \global\setbox\ZoneBBOX=\box\VOIDBOX
 \def\Zone{\InZoneB}
 \loop
  \ifMakePageInComplete
 \FindNext
 \ifnum\StackPointer=\m@ne
  \NextItem=\m@ne
  \MoreFiguresfalse
  \MoreTablesfalse
 \fi
 \ifnum\NextItem=\Figure
   \FindItem{\Figure}{\NextFigure}
   \ifnum\StackPointer=\m@ne \MoreFiguresfalse
   \else
    \GetItemSPAN{\StackPointer}
    \ifnum\ItemSPAN=\Double\relax
     \MoreFiguresfalse\fi
   \fi
   \ifMoreFigures\Print{}\FigureItems\fi
 \fi
 \ifnum\NextItem=\Table
   \FindItem{\Table}{\NextTable}
   \ifnum\StackPointer=\m@ne \MoreTablesfalse
   \else
    \GetItemSPAN{\StackPointer}
    \ifnum\ItemSPAN=\Double\relax
     \MoreTablesfalse\fi
   \fi
   \ifMoreTables\Print{}\TableItems\fi
 \fi
   \MakePageInCompletefalse 
   \ifMoreFigures\MakePageInCompletetrue\fi
   \ifMoreTables\MakePageInCompletetrue\fi
 \repeat
 \ifZoneAFullPage
  \global\TextSize=\z@
  \global\ZoneBSize=\z@
  \global\vsize=\z@\relax
  \global\topskip=\z@\relax
  \vbox to \z@{\vss}
  \eject
 \else
 \global\vsize=\ZoneBSize
 \global\hsize=\ColumnWidth
 \ifdim\TextSize<23pt
 \Warn{}
 \Warn{* Making column fall short: TextSize=\the\TextSize *}
 \vskip-\lastskip\eject\fi
\fi
}

\def\FigureItems{
 \Print{Considering...}
 \ShowItem{\StackPointer}
 \GetItemBOX{\StackPointer} 
 \GetItemSPAN{\StackPointer}
  \CheckFitInZone 
  \ifnum\ItemFits=\Yes
   \ifnum\ItemSPAN=\Single
     \ChangeStatus{\StackPointer}{\InZoneB} 
     \global\FigInZoneBtrue
     \ifFirstSingleItem
      \hbox{}\vskip-\BodgeHeight
     \global\advance\ItemSIZE by \TextLeading
     \fi
     \unvbox\ItemBOX\ItemSep
     \global\FirstSingleItemfalse
     \global\advance\TextSize by -\ItemSIZE
     \global\advance\TextSize by -\TextLeading
   \else
    \ifFirstZoneA
     \global\advance\ItemSIZE by \TextLeading
     \global\FirstZoneAfalse\fi
    \global\advance\TextSize by -\ItemSIZE
    \global\advance\TextSize by -\TextLeading
    \global\advance\ZoneBSize by -\ItemSIZE
    \global\advance\ZoneBSize by -\TextLeading
    \ifFigInZoneB\relax
     \else
     \ifdim\TextSize<3\TextLeading
     \global\ZoneAFullPagetrue
     \fi
    \fi
    \ChangeStatus{\StackPointer}{\Zone}
    \ifnum\Zone=\InZoneC \global\FigInZoneCtrue\fi
  \fi
   \Print{TextSize=\the\TextSize}
   \Print{ZoneBSize=\the\ZoneBSize}
  \global\advance\NextFigure \@ne
   \Print{This figure has been placed.}
  \else
   \Print{No space available for this figure...holding over.}
   \Print{}
   \global\MoreFiguresfalse
  \fi
}

\def\TableItems{
 \Print{Considering...}
 \ShowItem{\StackPointer}
 \GetItemBOX{\StackPointer} 
 \GetItemSPAN{\StackPointer}
  \CheckFitInZone 
  \ifnum\ItemFits=\Yes
   \ifnum\ItemSPAN=\Single
    \ChangeStatus{\StackPointer}{\InZoneB}
     \global\TabInZoneBtrue
     \ifFirstSingleItem
      \hbox{}\vskip-\BodgeHeight
     \global\advance\ItemSIZE by \TextLeading
     \fi
     \unvbox\ItemBOX\ItemSep
     \global\FirstSingleItemfalse
     \global\advance\TextSize by -\ItemSIZE
     \global\advance\TextSize by -\TextLeading
   \else
    \ifFirstZoneA
    \global\advance\ItemSIZE by \TextLeading
    \global\FirstZoneAfalse\fi
    \global\advance\TextSize by -\ItemSIZE
    \global\advance\TextSize by -\TextLeading
    \global\advance\ZoneBSize by -\ItemSIZE
    \global\advance\ZoneBSize by -\TextLeading
    \ifFigInZoneB\relax
     \else
     \ifdim\TextSize<3\TextLeading
     \global\ZoneAFullPagetrue
     \fi
    \fi
    \ChangeStatus{\StackPointer}{\Zone}
    \ifnum\Zone=\InZoneC \global\TabInZoneCtrue\fi
   \fi
  \global\advance\NextTable \@ne
   \Print{This table has been placed.}
  \else
  \Print{No space available for this table...holding over.}
   \Print{}
   \global\MoreTablesfalse
  \fi
}


\def\CheckFitInZone{%
{\advance\TextSize by -\ItemSIZE
 \advance\TextSize by -\TextLeading
 \ifFirstSingleItem
  \advance\TextSize by \TextLeading
 \fi
 \ifnum\Zone=\InZoneA\relax
  \else \advance\TextSize by -\ZoneBAdjust
 \fi
 \ifdim\TextSize<3\TextLeading \global\ItemFits=\No
 \else \global\ItemFits=\Yes\fi}
}

\def\BeginOpening{%
  \ninepoint
  \thispagestyle{titlepage}%
  \global\setbox\ItemBOX=\vbox\bgroup%
    \hsize=\PageWidth%
    \hrule height \z@
    \ifsinglecol\vskip 6pt\fi 
}

\let\begintopmatter=\BeginOpening  

\def\EndOpening{%
  \One
  \egroup
  \ifsinglecol
    \box\ItemBOX%
    \vskip\TextLeading plus 2\TextLeading
    \@noafterindent
  \else
    \ItemNUMBER=\z@%
    \ItemTYPE=\Figure
    \ItemSPAN=\Double
    \ItemSTATUS=\InStack
    \JoinStack
  \fi
}


\newif\if@here  \@herefalse

\def\no@float{\global\@heretrue}
\let\nofloat=\relax 

\def\beginfigure{%
  \@ifstar{\global\@dfloattrue \@bfigure}{\global\@dfloatfalse \@bfigure}%
}

\def\@bfigure#1{%
  \par
  \if@dfloat
    \ItemSPAN=\Double
    \TEMPDIMEN=\PageWidth
  \else
    \ItemSPAN=\Single
    \TEMPDIMEN=\ColumnWidth
  \fi
  \ifsinglecol
    \TEMPDIMEN=\PageWidth
  \else
    \ItemSTATUS=\InStack
    \ItemNUMBER=#1%
    \ItemTYPE=\Figure
  \fi
  \bgroup
    \hsize=\TEMPDIMEN
    \global\setbox\ItemBOX=\vbox\bgroup
      \eightpoint\nostb@ls{10pt}%
      \let\caption=\fig@caption
      \ifsinglecol \let\nofloat=\no@float\fi
}

\def\fig@caption#1{%
  \vskip 5.5pt plus 6pt%
  \bgroup 
    \eightpoint\nostb@ls{10pt}%
    \setbox\TEMPBOX=\hbox{#1}%
    \ifdim\wd\TEMPBOX>\TEMPDIMEN
      \noindent \unhbox\TEMPBOX\par
    \else
      \hbox to \hsize{\hfil\unhbox\TEMPBOX\hfil}%
    \fi
  \egroup
}

\def\endfigure{%
  \par\egroup 
  \egroup
  \ifsinglecol
    \if@here \midinsert\global\@herefalse\else \topinsert\fi
      \unvbox\ItemBOX
    \endinsert
  \else
    \JoinStack
    \Print{Processing source for figure \the\ItemNUMBER}%
  \fi
}


\newbox\tab@cap@box
\def\tab@caption#1{\global\setbox\tab@cap@box=\hbox{#1\par}}

\newtoks\tab@txt@toks
\long\def\tab@txt#1{\global\tab@txt@toks={#1}\global\table@txttrue}

\newif\iftable@txt  \table@txtfalse
\newif\if@dfloat    \@dfloatfalse

\def\begintable{%
  \@ifstar{\global\@dfloattrue \@btable}{\global\@dfloatfalse \@btable}%
}

\def\@btable#1{%
  \par
  \if@dfloat
    \ItemSPAN=\Double
    \TEMPDIMEN=\PageWidth
  \else
    \ItemSPAN=\Single
    \TEMPDIMEN=\ColumnWidth
  \fi
  \ifsinglecol
    \TEMPDIMEN=\PageWidth
  \else
    \ItemSTATUS=\InStack
    \ItemNUMBER=#1%
    \ItemTYPE=\Table
  \fi
  \bgroup
    \eightpoint\nostb@ls{10pt}%
    \global\setbox\ItemBOX=\vbox\bgroup
      \let\caption=\tab@caption
      \let\tabletext=\tab@txt
      \ifsinglecol \let\nofloat=\no@float\fi
}

\def\endtable{%
  \par\egroup 
  \egroup
  \setbox\TEMPBOX=\hbox to \TEMPDIMEN{%
    \eightpoint\nostb@ls{10pt}%
    \hss
    \vbox{%
      \hsize=\wd\ItemBOX
      \ifvoid\tab@cap@box
      \else
        \noindent\unhbox\tab@cap@box
        \vskip 5.5pt plus 6pt%
      \fi
      \box\ItemBOX
      \iftable@txt
        \vskip 10pt%
        \noindent\the\tab@txt@toks
        \global\table@txtfalse
      \fi
    }%
    \hss
  }%
  \ifsinglecol
    \if@here \midinsert\global\@herefalse\else \topinsert\fi
      \box\TEMPBOX
    \endinsert
  \else
    \global\setbox\ItemBOX=\box\TEMPBOX
    \JoinStack
    \Print{Processing source for table \the\ItemNUMBER}%
  \fi
}

\def\UnloadZoneA{%
\FirstZoneAtrue
 \Iteration=\z@
  \loop
   \ifnum\Iteration<\LengthOfStack
    \GetItemSTATUS{\Iteration}
    \ifnum\ItemSTATUS=\InZoneA
     \GetItemBOX{\Iteration}
     \ifFirstZoneA \vbox to \BodgeHeight{\vfil}%
     \FirstZoneAfalse\fi
     \unvbox\ItemBOX\ItemSep
     \LeaveStack{\Iteration}
     \else
     \advance\Iteration \@ne
   \fi
 \repeat
}

\def\UnloadZoneC{%
\Iteration=\z@
  \loop
   \ifnum\Iteration<\LengthOfStack
    \GetItemSTATUS{\Iteration}
    \ifnum\ItemSTATUS=\InZoneC
     \GetItemBOX{\Iteration}
     \ItemSep\unvbox\ItemBOX
     \LeaveStack{\Iteration}
     \else
     \advance\Iteration \@ne
   \fi
 \repeat
}


\def\ShowItem#1{
  {\GetItemAll{#1}
  \Print{\the#1:
  {TYPE=\ifnum\ItemTYPE=\Figure Figure\else Table\fi}
  {NUMBER=\the\ItemNUMBER}
  {SPAN=\ifnum\ItemSPAN=\Single Single\else Double\fi}
  {SIZE=\the\ItemSIZE}}}
}

\def\ShowStack{%
 \Print{}
 \Print{LengthOfStack = \the\LengthOfStack}
 \ifnum\LengthOfStack=\z@ \Print{Stack is empty}\fi
 \Iteration=\z@
 \loop
 \ifnum\Iteration<\LengthOfStack
  \ShowItem{\Iteration}
  \advance\Iteration \@ne
 \repeat
}

\def\B#1#2{%
\hbox{\vrule\kern-0.4pt\vbox to #2{%
\hrule width #1\vfill\hrule}\kern-0.4pt\vrule}
}


\newif\ifsinglecol   \singlecolfalse

\def\onecolumn{%
  \global\output={\singlecoloutput}%
  \global\hsize=\PageWidth
  \global\vsize=\PageHeight
  \global\ColumnWidth=\hsize
  \global\TextLeading=12pt
  \global\Leading=12
  \global\singlecoltrue
  \global\let\onecolumn=\relax
  \global\let\footnote=\sing@footnote
  \global\let\vfootnote=\sing@vfootnote
  \ninepoint 
  \message{(Single column)}%
}

\def\singlecoloutput{%
  \shipout\vbox{\PageHead\vbox to \PageHeight{\pagebody\vss}\PageFoot}%
  \advancepageno
  \ifplate@page
    \shipout\vbox{%
      \sp@pagetrue
      \def\sp@type{plate}%
      \global\plate@pagefalse
      \PageHead\vbox to \PageHeight{\unvbox\plt@box\vfil}\PageFoot%
    }%
    \message{[plate]}%
    \advancepageno
  \fi
  \ifnum\outputpenalty>-\@MM \else\dosupereject\fi%
}

\def\ItemSep{\vskip\ItemSepamount\relax}

\def\ItemSepbreak{\par\ifdim\lastskip<\ItemSepamount
  \removelastskip\penalty-200\ItemSep\fi%
}


\let\@@endinsert=\endinsert 

\def\endinsert{\egroup 
  \if@mid \dimen@\ht\z@ \advance\dimen@\dp\z@ \advance\dimen@12\p@
    \advance\dimen@\pagetotal \advance\dimen@-\pageshrink
    \ifdim\dimen@>\pagegoal\@midfalse\p@gefalse\fi\fi
  \if@mid \ItemSep\box\z@\ItemSepbreak
  \else\insert\topins{\penalty100 
    \splittopskip\z@skip
    \splitmaxdepth\maxdimen \floatingpenalty\z@
    \ifp@ge \dimen@\dp\z@
    \vbox to\vsize{\unvbox\z@\kern-\dimen@}
    \else \box\z@\nobreak\ItemSep\fi}\fi\endgroup%
}


\def\gobbleone#1{}
\def\gobbletwo#1#2{}
\let\footnote=\gobbletwo 
\let\vfootnote=\gobbleone

\def\sing@footnote#1{\let\@sf\empty 
  \ifhmode\edef\@sf{\spacefactor\the\spacefactor}\/\fi
  \hbox{$^{\hbox{\eightpoint #1}}$}\@sf\sing@vfootnote{#1}%
}

\def\sing@vfootnote#1{\insert\footins\bgroup\eightpoint\b@ls{9pt}%
  \interlinepenalty\interfootnotelinepenalty
  \splittopskip\ht\strutbox 
  \splitmaxdepth\dp\strutbox \floatingpenalty\@MM
  \leftskip\z@skip \rightskip\z@skip \spaceskip\z@skip \xspaceskip\z@skip
  \noindent $^{\scriptstyle\hbox{#1}}$\hskip 4pt%
    \footstrut\futurelet\next\fo@t%
}

\def\footnoterule{\kern-3\p@ \hrule height \z@ \kern 3\p@}

\skip\footins=19.5pt plus 12pt minus 1pt
\count\footins=1000
\dimen\footins=\maxdimen

\def\note#1#2{%
  \let\@sf=\empty \ifhmode\edef\@sf{\spacefactor\the\spacefactor}\/\fi
  #1\insert\footins\bgroup
    \eightpoint\b@ls{10pt}\rm
    \interlinepenalty\interfootnotelinepenalty
    \splitmaxdepth\dp\strutbox \floatingpenalty\@MM
    \leftskip\z@skip \rightskip\z@skip \spaceskip\z@skip \xspaceskip\z@skip
    \noindent\footstrut #1$\,$#2\strut\par
  \egroup
  \@sf\relax}


\def\landscape{%
  \global\TEMPDIMEN=\PageWidth
  \global\PageWidth=\PageHeight
  \global\PageHeight=\TEMPDIMEN
  \global\let\landscape=\relax
  \onecolumn
  \message{(landscape)}%
  \raggedbottom
}


\output{%
  \ifLeftCOL
    \global\setbox\LeftBOX=\vbox to \ZoneBSize{\box255\unvbox\ZoneBBOX
      \ifvoid\footins\else
        \vskip\skip\footins\unvbox\footins\fi
    }%
    \global\LeftCOLfalse
    \MakeRightCol
  \else
    \setbox\RightBOX=\vbox to \ZoneBSize{\box255\unvbox\ZoneBBOX
      \ifvoid\footins\else
        \vskip\skip\footins\unvbox\footins\fi
    }%
    \setbox\MidBOX=\hbox{\box\LeftBOX\hskip\ColumnGap\box\RightBOX}%
    \setbox\PageBOX=\vbox to \PageHeight{%
      \UnloadZoneA\box\MidBOX\UnloadZoneC}%
    \shipout\vbox{\PageHead\vbox to \PageHeight{\box\PageBOX\vss}\PageFoot}%
    \advancepageno
    \ifplate@page
      \shipout\vbox{%
        \sp@pagetrue
        \def\sp@type{plate}%
        \global\plate@pagefalse
        \PageHead\vbox to \PageHeight{\unvbox\plt@box\vfil}\PageFoot%
      }%
      \message{[plate]}%
      \advancepageno
    \fi
    \global\LeftCOLtrue
    \CleanStack
    \MakePage
  \fi
}


\Warn{\start@mess}

\newif\ifCUPmtplainloaded 
\ifprod@font
  \global\CUPmtplainloadedtrue
\fi

\def\mnmacrosloaded{} 

\catcode `\@=12 



\fi

%

\newif\ifAMStwofonts

\ifCUPmtplainloaded \else
  \NewTextAlphabet{textbfit} {cmbxti10} {}
  \NewTextAlphabet{textbfss} {cmssbx10} {}
  \NewMathAlphabet{mathbfit} {cmbxti10} {} 
  \NewMathAlphabet{mathbfss} {cmssbx10} {} 
  \ifAMStwofonts
    \NewSymbolFont{upmath} {eurm10}
    \NewSymbolFont{AMSa} {msam10}
    \NewMathSymbol{\upi}     {0}{upmath}{19}
    \NewMathSymbol{\umu}     {0}{upmath}{16}
    \NewMathSymbol{\upartial}{0}{upmath}{40}
    \NewMathSymbol{\leqslant}{3}{AMSa}{36}
    \NewMathSymbol{\geqslant}{3}{AMSa}{3E}

    \let\leq=\leqslant \let\le=\leqslant
    \let\geq=\geqslant \let\ge=\geqslant
  \else
    \def\umu{\mu}
    \def\upi{\pi}
    \def\upartial{\partial}
  \fi
\fi


\pageoffset{-2.5pc}{0pc}

\loadboldmathnames



\pagerange{0--0}    
\pubyear{0000}
\volume{000}

\newcount\eqcount\eqcount=0
\newcount\figcount\figcount=0
\def\nextenum{\global\advance\eqcount by 1 \thenumber}
\def\nextfignum{\global\advance\figcount by 1 \thefignumber}
\def\thenumber{\the\eqcount}
\def\thefignumber{\the\figcount}
\def\nameeq#1{\xdef#1{\thenumber}\ignorespaces}
\def\namefig#1{\xdef#1{\thefignumber}\ignorespaces}
\def\n{\nextenum}
\def\nf{\nextfignum}
\def\na#1{\label{2}{\string#1}\nameeq#1}
\def\naf#1{\label{2}{\string#1}\namefig#1}

\newbox\labelbox\newdimen\prevprevdepth
\def\label#1#2{%
\setbox\labelbox%
\vbox to 0pt{\kern -#1\baselineskip%
\hbox to 0pt{\kern\hsize\kern1pt\fivei#2\hss}%
\vss}%
\ifvmode\prevprevdepth=\prevdepth
\vskip-\baselineskip\vskip-\parskip%
\vskip\prevprevdepth%
\box\labelbox%
\prevdepth=\prevprevdepth
\else\vadjust{\box\labelbox}\fi%
}

\def\offlabel{\def\label##1##2{}}

\def\Msun{{\ifmmode M_{\sun} \else $M_{\sun}$ \fi}}
\def\msun{{\ifmmode m_{\sun} \else $m_{\sun}$ \fi}}
\def\Lsun{{\ifmmode L_{\sun} \else $L_{\sun}$ \fi}}
\def\rsun{{\ifmmode r_{\sun} \else $r_{\sun}$ \fi}}
\def\Rsun{{\ifmmode R_{\sun} \else $R_{\sun}$ \fi}}

\def\la{\mathrel{\mathchoice {\vcenter{\offinterlineskip\halign{\hfil
$\displaystyle##$\hfil\cr<\cr\sim\cr}}}
{\vcenter{\offinterlineskip\halign{\hfil$\textstyle##$\hfil\cr<\cr\sim\cr}}}
{\vcenter{\offinterlineskip\halign{\hfil$\scriptstyle##$\hfil\cr<\cr\sim\cr}}}
{\vcenter{\offinterlineskip\halign{\hfil$\scriptscriptstyle##$\hfil\cr<\cr\sim
\cr}}}}}
\def\ga{\mathrel{\mathchoice {\vcenter{\offinterlineskip\halign{\hfil
$\displaystyle##$\hfil\cr>\cr\sim\cr}}}
{\vcenter{\offinterlineskip\halign{\hfil$\textstyle##$\hfil\cr>\cr\sim\cr}}}
{\vcenter{\offinterlineskip\halign{\hfil$\scriptstyle##$\hfil\cr>\cr\sim\cr}}}
{\vcenter{\offinterlineskip\halign{\hfil$\scriptscriptstyle##$\hfil\cr>\cr\sim
\cr}}}}}

\def\PsfigVersion{1.10}
\def\setDriver{\DvipsDriver} 
\ifx\undefined\psfig\else \fi
%

\let\LaTeXAtSign=\@
\let\@=\relax
\edef\psfigRestoreAt{\catcode`\@=\number\catcode`@\relax}
\catcode`\@=11\relax
\newwrite\@unused
\def\ps@typeout#1{{\let\protect\string\immediate\write\@unused{#1}}}

\def\DvipsDriver{
	\ps@typeout{psfig/tex \PsfigVersion -dvips}
\def\PsfigSpecials{\DvipsSpecials} 	\def\ps@dir{/}
\def\ps@predir{} }
\def\OzTeXDriver{
	\ps@typeout{psfig/tex \PsfigVersion -oztex}
	\def\PsfigSpecials{\OzTeXSpecials}
	\def\ps@dir{:}
	\def\ps@predir{:}
	\catcode`\^^J=5
}


\def\figurepath{./:}

\def\DoPaths#1{\expandafter\EachPath#1\stoplist}
\def\leer{}
\def\EachPath#1:#2\stoplist{
  \ExistsFile{#1}{\SearchedFile}
  \ifx#2\leer
  \else
    \expandafter\EachPath#2\stoplist
  \fi}
%
%
\def\ps@dir{/}
\def\ExistsFile#1#2{%
   \openin1=\ps@predir#1\ps@dir#2
   \ifeof1
       \closein1
   \else
       \closein1
        \ifx\ps@founddir\leer
           \edef\ps@founddir{#1}
        \fi
   \fi}
%
%
\def\get@dir#1{%
  \def\ps@founddir{}
  \def\SearchedFile{#1}
  \DoPaths\figurepath
}

%
%
\def\@nnil{\@nil}
\def\@empty{}
\def\@psdonoop#1\@@#2#3{}
\def\@psdo#1:=#2\do#3{\edef\@psdotmp{#2}\ifx\@psdotmp\@empty \else
    \expandafter\@psdoloop#2,\@nil,\@nil\@@#1{#3}\fi}
\def\@psdoloop#1,#2,#3\@@#4#5{\def#4{#1}\ifx #4\@nnil \else
       #5\def#4{#2}\ifx #4\@nnil \else#5\@ipsdoloop #3\@@#4{#5}\fi\fi}
\def\@ipsdoloop#1,#2\@@#3#4{\def#3{#1}\ifx #3\@nnil 
       \let\@nextwhile=\@psdonoop \else
      #4\relax\let\@nextwhile=\@ipsdoloop\fi\@nextwhile#2\@@#3{#4}}
\def\@tpsdo#1:=#2\do#3{\xdef\@psdotmp{#2}\ifx\@psdotmp\@empty \else
    \@tpsdoloop#2\@nil\@nil\@@#1{#3}\fi}
\def\@tpsdoloop#1#2\@@#3#4{\def#3{#1}\ifx #3\@nnil 
       \let\@nextwhile=\@psdonoop \else
      #4\relax\let\@nextwhile=\@tpsdoloop\fi\@nextwhile#2\@@#3{#4}}
%
\ifx\undefined\fbox
\newdimen\fboxrule
\newdimen\fboxsep
\newdimen\ps@tempdima
\newbox\ps@tempboxa
\fboxsep = 3pt
\fboxrule = .4pt
\long\def\fbox#1{\leavevmode\setbox\ps@tempboxa\hbox{#1}\ps@tempdima\fboxrule
    \advance\ps@tempdima \fboxsep \advance\ps@tempdima \dp\ps@tempboxa
   \hbox{\lower \ps@tempdima\hbox
  {\vbox{\hrule height \fboxrule
          \hbox{\vrule width \fboxrule \hskip\fboxsep
          \vbox{\vskip\fboxsep \box\ps@tempboxa\vskip\fboxsep}\hskip 
                 \fboxsep\vrule width \fboxrule}
                 \hrule height \fboxrule}}}}
\fi
%
%
\newread\ps@stream
\newif\ifnot@eof       
\newif\if@noisy        
\newif\if@atend        
\newif\if@psfile       
%
%
{\catcode`\%=12\global\gdef\epsf@start{
\def\epsf@PS{PS}
\def\epsf@getbb#1{%
%
%
\openin\ps@stream=\ps@predir#1
\ifeof\ps@stream\ps@typeout{Error, File #1 not found}\else
%
%
   {\not@eoftrue \chardef\other=12
    \def\do##1{\catcode`##1=\other}\dospecials \catcode`\ =10
    \loop
       \if@psfile
	  \read\ps@stream to \epsf@fileline
       \else{
	  \obeyspaces
          \read\ps@stream to \epsf@tmp\global\let\epsf@fileline\epsf@tmp}
       \fi
       \ifeof\ps@stream\not@eoffalse\else
%
%
       \if@psfile\else
       \expandafter\epsf@test\epsf@fileline:. \\%
       \fi
%
%
          \expandafter\epsf@aux\epsf@fileline:. \\%
       \fi
   \ifnot@eof\repeat
   }\closein\ps@stream\fi}%
%
%
\long\def\epsf@test#1#2#3:#4\\{\def\epsf@testit{#1#2}
			\ifx\epsf@testit\epsf@start\else
\ps@typeout{Warning! File does not start with `\epsf@start'.  It may not be a PostScript file.}
			\fi
			\@psfiletrue} 
%
%
{\catcode`\%=12\global\let\epsf@percent=
%
%
%
\long\def\epsf@aux#1#2:#3\\{\ifx#1\epsf@percent
   \def\epsf@testit{#2}\ifx\epsf@testit\epsf@bblit
	\@atendfalse
        \epsf@atend #3 . \\%
	\if@atend	
	   \if@verbose{
		\ps@typeout{psfig: found `(atend)'; continuing search}
	   }\fi
        \else
        \epsf@grab #3 . . . \\%
        \not@eoffalse
        \global\no@bbfalse
        \fi
   \fi\fi}%
%
%
\def\epsf@grab #1 #2 #3 #4 #5\\{%
   \global\def\epsf@llx{#1}\ifx\epsf@llx\empty
      \epsf@grab #2 #3 #4 #5 .\\\else
   \global\def\epsf@lly{#2}%
   \global\def\epsf@urx{#3}\global\def\epsf@ury{#4}\fi}%
%
%
\def\epsf@atendlit{(atend)} 
\def\epsf@atend #1 #2 #3\\{%
   \def\epsf@tmp{#1}\ifx\epsf@tmp\empty
      \epsf@atend #2 #3 .\\\else
   \ifx\epsf@tmp\epsf@atendlit\@atendtrue\fi\fi}


\chardef\psletter = 11 
\chardef\other = 12

\newif \ifdebug 
\newif\ifc@mpute 
\c@mputetrue 

\let\then = \relax
\def\r@dian{pt }
\let\r@dians = \r@dian
\let\dimensionless@nit = \r@dian
\let\dimensionless@nits = \dimensionless@nit
\def\internal@nit{sp }
\let\internal@nits = \internal@nit
\newif\ifstillc@nverging
\def \Mess@ge #1{\ifdebug \then \message {#1} \fi}

{ 
	\catcode `\@ = \psletter
	\gdef \nodimen {\expandafter \n@dimen \the \dimen}
	\gdef \term #1 #2 #3%
	       {\edef \t@ {\the #1}
		\edef \t@@ {\expandafter \n@dimen \the #2\r@dian}%
		\t@rm {\t@} {\t@@} {#3}%
	       }
	\gdef \t@rm #1 #2 #3%
	       {{%
		\count 0 = 0
		\dimen 0 = 1 \dimensionless@nit
		\dimen 2 = #2\relax
		\Mess@ge {Calculating term #1 of \nodimen 2}%
		\loop
		\ifnum	\count 0 < #1
		\then	\advance \count 0 by 1
			\Mess@ge {Iteration \the \count 0 \space}%
			\Multiply \dimen 0 by {\dimen 2}%
			\Mess@ge {After multiplication, term = \nodimen 0}%
			\Divide \dimen 0 by {\count 0}%
			\Mess@ge {After division, term = \nodimen 0}%
		\repeat
		\Mess@ge {Final value for term #1 of 
				\nodimen 2 \space is \nodimen 0}%
		\xdef \Term {#3 = \nodimen 0 \r@dians}%
		\aftergroup \Term
	       }}
	\catcode `\p = \other
	\catcode `\t = \other
	\gdef \n@dimen #1pt{#1} 
}

\def \Divide #1by #2{\divide #1 by #2} 

\def \Multiply #1by #2
       {{
	\count 0 = #1\relax
	\count 2 = #2\relax
	\count 4 = 65536
	\Mess@ge {Before scaling, count 0 = \the \count 0 \space and
			count 2 = \the \count 2}%
	\ifnum	\count 0 > 32767 
	\then	\divide \count 0 by 4
		\divide \count 4 by 4
	\else	\ifnum	\count 0 < -32767
		\then	\divide \count 0 by 4
			\divide \count 4 by 4
		\else
		\fi
	\fi
	\ifnum	\count 2 > 32767 
	\then	\divide \count 2 by 4
		\divide \count 4 by 4
	\else	\ifnum	\count 2 < -32767
		\then	\divide \count 2 by 4
			\divide \count 4 by 4
		\else
		\fi
	\fi
	\multiply \count 0 by \count 2
	\divide \count 0 by \count 4
	\xdef \product {#1 = \the \count 0 \internal@nits}%
	\aftergroup \product
       }}

\def\r@duce{\ifdim\dimen0 > 90\r@dian \then   
		\multiply\dimen0 by -1
		\advance\dimen0 by 180\r@dian
		\r@duce
	    \else \ifdim\dimen0 < -90\r@dian \then  
		\advance\dimen0 by 360\r@dian
		\r@duce
		\fi
	    \fi}

\def\Sine#1%
       {{%
	\dimen 0 = #1 \r@dian
	\r@duce
	\ifdim\dimen0 = -90\r@dian \then
	   \dimen4 = -1\r@dian
	   \c@mputefalse
	\fi
	\ifdim\dimen0 = 90\r@dian \then
	   \dimen4 = 1\r@dian
	   \c@mputefalse
	\fi
	\ifdim\dimen0 = 0\r@dian \then
	   \dimen4 = 0\r@dian
	   \c@mputefalse
	\fi
	\ifc@mpute \then
		\divide\dimen0 by 180
		\dimen0=3.141592654\dimen0
		\dimen 2 = 3.1415926535897963\r@dian 
		\divide\dimen 2 by 2 
		\Mess@ge {Sin: calculating Sin of \nodimen 0}%
		\count 0 = 1 
		\dimen 2 = 1 \r@dian 
		\dimen 4 = 0 \r@dian 
		\loop
			\ifnum	\dimen 2 = 0 
			\then	\stillc@nvergingfalse 
			\else	\stillc@nvergingtrue
			\fi
			\ifstillc@nverging 
			\then	\term {\count 0} {\dimen 0} {\dimen 2}%
				\advance \count 0 by 2
				\count 2 = \count 0
				\divide \count 2 by 2
				\ifodd	\count 2 
				\then	\advance \dimen 4 by \dimen 2
				\else	\advance \dimen 4 by -\dimen 2
				\fi
		\repeat
	\fi		
			\xdef \sine {\nodimen 4}%
       }}

\def\Cosine#1{\ifx\sine\UnDefined\edef\Savesine{\relax}\else
		             \edef\Savesine{\sine}\fi
	{\dimen0=#1\r@dian\advance\dimen0 by 90\r@dian
	 \Sine{\nodimen 0}
	 \xdef\cosine{\sine}
	 \xdef\sine{\Savesine}}}	      

\def\psdraft{
	\def\@psdraft{0}
}
\def\psfull{
	\def\@psdraft{100}
}

\psfull

\newif\if@scalefirst
\def\psscalefirst{\@scalefirsttrue}
\def\psrotatefirst{\@scalefirstfalse}
\psrotatefirst

\newif\if@draftbox
\def\psnodraftbox{
	\@draftboxfalse
}
\def\psdraftbox{
	\@draftboxtrue
}
\@draftboxtrue

\newif\if@prologfile
\newif\if@postlogfile
\def\pssilent{
	\@noisyfalse
}
\def\psnoisy{
	\@noisytrue
}
\psnoisy
\newif\if@bbllx
\newif\if@bblly
\newif\if@bburx
\newif\if@bbury
\newif\if@height
\newif\if@width
\newif\if@rheight
\newif\if@rwidth
\newif\if@angle
\newif\if@clip
\newif\if@verbose
\def\@p@@sclip#1{\@cliptrue}
\newif\if@decmpr
\def\@p@@sfigure#1{\def\@p@sfile{null}\def\@p@sbbfile{null}\@decmprfalse
   \openin1=\ps@predir#1
   \ifeof1
	\closein1
	\get@dir{#1}
	\ifx\ps@founddir\leer
		\openin1=\ps@predir#1.bb
		\ifeof1
			\closein1
			\get@dir{#1.bb}
			\ifx\ps@founddir\leer
				\ps@typeout{Can't find #1 in \figurepath}
			\else
				\@decmprtrue
				\def\@p@sfile{\ps@founddir\ps@dir#1}
				\def\@p@sbbfile{\ps@founddir\ps@dir#1.bb}
			\fi
		\else
			\closein1
			\@decmprtrue
			\def\@p@sfile{#1}
			\def\@p@sbbfile{#1.bb}
		\fi
	\else
		\def\@p@sfile{\ps@founddir\ps@dir#1}
		\def\@p@sbbfile{\ps@founddir\ps@dir#1}
	\fi
   \else
	\closein1
	\def\@p@sfile{#1}
	\def\@p@sbbfile{#1}
   \fi
}
\def\@p@@sfile#1{\@p@@sfigure{#1}}
\def\@p@@sbbllx#1{
		\@bbllxtrue
		\dimen100=#1
		\edef\@p@sbbllx{\number\dimen100}
}
\def\@p@@sbblly#1{
		\@bbllytrue
		\dimen100=#1
		\edef\@p@sbblly{\number\dimen100}
}
\def\@p@@sbburx#1{
		\@bburxtrue
		\dimen100=#1
		\edef\@p@sbburx{\number\dimen100}
}
\def\@p@@sbbury#1{
		\@bburytrue
		\dimen100=#1
		\edef\@p@sbbury{\number\dimen100}
}
\def\@p@@sheight#1{
		\@heighttrue
		\dimen100=#1
   		\edef\@p@sheight{\number\dimen100}
}
\def\@p@@swidth#1{
		\@widthtrue
		\dimen100=#1
		\edef\@p@swidth{\number\dimen100}
}
\def\@p@@srheight#1{
		\@rheighttrue
		\dimen100=#1
		\edef\@p@srheight{\number\dimen100}
}
\def\@p@@srwidth#1{
		\@rwidthtrue
		\dimen100=#1
		\edef\@p@srwidth{\number\dimen100}
}
\def\@p@@sangle#1{
		\@angletrue
		\edef\@p@sangle{#1} 
}
\def\@p@@ssilent#1{ 
		\@verbosefalse
}
\def\@p@@sprolog#1{\@prologfiletrue\def\@prologfileval{#1}}
\def\@p@@spostlog#1{\@postlogfiletrue\def\@postlogfileval{#1}}
\def\@cs@name#1{\csname #1\endcsname}
\def\@setparms#1=#2,{\@cs@name{@p@@s#1}{#2}}
%
%
\def\ps@init@parms{
		\@bbllxfalse \@bbllyfalse
		\@bburxfalse \@bburyfalse
		\@heightfalse \@widthfalse
		\@rheightfalse \@rwidthfalse
		\def\@p@sbbllx{}\def\@p@sbblly{}
		\def\@p@sbburx{}\def\@p@sbbury{}
		\def\@p@sheight{}\def\@p@swidth{}
		\def\@p@srheight{}\def\@p@srwidth{}
		\def\@p@sangle{0}
		\def\@p@sfile{} \def\@p@sbbfile{}
		\def\@p@scost{10}
		\def\@sc{}
		\@prologfilefalse
		\@postlogfilefalse
		\@clipfalse
		\if@noisy
			\@verbosetrue
		\else
			\@verbosefalse
		\fi
}
%
%
\def\parse@ps@parms#1{
	 	\@psdo\@psfiga:=#1\do
		   {\expandafter\@setparms\@psfiga,}}
%
%
\newif\ifno@bb
\def\bb@missing{
	\if@verbose{
		\ps@typeout{psfig: searching \@p@sbbfile \space  for bounding box}
	}\fi
	\no@bbtrue
	\epsf@getbb{\@p@sbbfile}
        \ifno@bb \else \bb@cull\epsf@llx\epsf@lly\epsf@urx\epsf@ury\fi
}	
\def\bb@cull#1#2#3#4{
	\dimen100=#1 bp\edef\@p@sbbllx{\number\dimen100}
	\dimen100=#2 bp\edef\@p@sbblly{\number\dimen100}
	\dimen100=#3 bp\edef\@p@sbburx{\number\dimen100}
	\dimen100=#4 bp\edef\@p@sbbury{\number\dimen100}
	\no@bbfalse
}
\newdimen\p@intvaluex
\newdimen\p@intvaluey
\def\rotate@#1#2{{\dimen0=#1 sp\dimen1=#2 sp
		  \global\p@intvaluex=\cosine\dimen0
		  \dimen3=\sine\dimen1
		  \global\advance\p@intvaluex by -\dimen3
		  \global\p@intvaluey=\sine\dimen0
		  \dimen3=\cosine\dimen1
		  \global\advance\p@intvaluey by \dimen3
		  }}
\def\compute@bb{
		\no@bbfalse
		\if@bbllx \else \no@bbtrue \fi
		\if@bblly \else \no@bbtrue \fi
		\if@bburx \else \no@bbtrue \fi
		\if@bbury \else \no@bbtrue \fi
		\ifno@bb \bb@missing \fi
		\ifno@bb \ps@typeout{FATAL ERROR: no bb supplied or found}
			\no-bb-error
		\fi
		%
%
		\count203=\@p@sbburx
		\count204=\@p@sbbury
		\advance\count203 by -\@p@sbbllx
		\advance\count204 by -\@p@sbblly
		\edef\ps@bbw{\number\count203}
		\edef\ps@bbh{\number\count204}
		\if@angle 
			\Sine{\@p@sangle}\Cosine{\@p@sangle}
	        	{\dimen100=\maxdimen\xdef\r@p@sbbllx{\number\dimen100}
					    \xdef\r@p@sbblly{\number\dimen100}
			                    \xdef\r@p@sbburx{-\number\dimen100}
					    \xdef\r@p@sbbury{-\number\dimen100}}
%
                        \def\minmaxtest{
			   \ifnum\number\p@intvaluex<\r@p@sbbllx
			      \xdef\r@p@sbbllx{\number\p@intvaluex}\fi
			   \ifnum\number\p@intvaluex>\r@p@sbburx
			      \xdef\r@p@sbburx{\number\p@intvaluex}\fi
			   \ifnum\number\p@intvaluey<\r@p@sbblly
			      \xdef\r@p@sbblly{\number\p@intvaluey}\fi
			   \ifnum\number\p@intvaluey>\r@p@sbbury
			      \xdef\r@p@sbbury{\number\p@intvaluey}\fi
			   }
			\rotate@{\@p@sbbllx}{\@p@sbblly}
			\minmaxtest
			\rotate@{\@p@sbbllx}{\@p@sbbury}
			\minmaxtest
			\rotate@{\@p@sbburx}{\@p@sbblly}
			\minmaxtest
			\rotate@{\@p@sbburx}{\@p@sbbury}
			\minmaxtest
			\edef\@p@sbbllx{\r@p@sbbllx}\edef\@p@sbblly{\r@p@sbblly}
			\edef\@p@sbburx{\r@p@sbburx}\edef\@p@sbbury{\r@p@sbbury}
		\fi
		\count203=\@p@sbburx
		\count204=\@p@sbbury
		\advance\count203 by -\@p@sbbllx
		\advance\count204 by -\@p@sbblly
		\edef\@bbw{\number\count203}
		\edef\@bbh{\number\count204}
}
%
%
\def\in@hundreds#1#2#3{\count240=#2 \count241=#3
		     \count100=\count240	
		     \divide\count100 by \count241
		     \count101=\count100
		     \multiply\count101 by \count241
		     \advance\count240 by -\count101
		     \multiply\count240 by 10
		     \count101=\count240	
		     \divide\count101 by \count241
		     \count102=\count101
		     \multiply\count102 by \count241
		     \advance\count240 by -\count102
		     \multiply\count240 by 10
		     \count102=\count240	
		     \divide\count102 by \count241
		     \count200=#1\count205=0
		     \count201=\count200
			\multiply\count201 by \count100
		 	\advance\count205 by \count201
		     \count201=\count200
			\divide\count201 by 10
			\multiply\count201 by \count101
			\advance\count205 by \count201
		     \count201=\count200
			\divide\count201 by 100
			\multiply\count201 by \count102
			\advance\count205 by \count201
		     \edef\@result{\number\count205}
}
\def\compute@wfromh{
		\in@hundreds{\@p@sheight}{\@bbw}{\@bbh}
		\edef\@p@swidth{\@result}
}
\def\compute@hfromw{
	        \in@hundreds{\@p@swidth}{\@bbh}{\@bbw}
		\edef\@p@sheight{\@result}
}
\def\compute@handw{
		\if@height 
			\if@width
			\else
				\compute@wfromh
			\fi
		\else 
			\if@width
				\compute@hfromw
			\else
				\edef\@p@sheight{\@bbh}
				\edef\@p@swidth{\@bbw}
			\fi
		\fi
}
\def\compute@resv{
		\if@rheight \else \edef\@p@srheight{\@p@sheight} \fi
		\if@rwidth \else \edef\@p@srwidth{\@p@swidth} \fi
}
%
\def\compute@sizes{
	\compute@bb
	\if@scalefirst\if@angle
	\if@width
	   \in@hundreds{\@p@swidth}{\@bbw}{\ps@bbw}
	   \edef\@p@swidth{\@result}
	\fi
	\if@height
	   \in@hundreds{\@p@sheight}{\@bbh}{\ps@bbh}
	   \edef\@p@sheight{\@result}
	\fi
	\fi\fi
	\compute@handw
	\compute@resv}
\def\OzTeXSpecials{
	\special{empty.ps /@isp {true} def}
	\special{empty.ps \@p@swidth \space \@p@sheight \space
			\@p@sbbllx \space \@p@sbblly \space
			\@p@sbburx \space \@p@sbbury \space
			startTexFig \space }
	\if@clip{
		\if@verbose{
			\ps@typeout{(clip)}
		}\fi
		\special{empty.ps doclip \space }
	}\fi
	\if@angle{
		\if@verbose{
			\ps@typeout{(rotate)}
		}\fi
		\special {empty.ps \@p@sangle \space rotate \space} 
	}\fi
	\if@prologfile
	    \special{\@prologfileval \space } \fi
	\if@decmpr{
		\if@verbose{
			\ps@typeout{psfig: Compression not available
			in OzTeX version \space }
		}\fi
	}\else{
		\if@verbose{
			\ps@typeout{psfig: including \@p@sfile \space }
		}\fi
		\special{epsf=\ps@predir\@p@sfile \space }
	}\fi
	\if@postlogfile
	    \special{\@postlogfileval \space } \fi
	\special{empty.ps /@isp {false} def}
}
\def\DvipsSpecials{
	\special{ps::[begin] 	\@p@swidth \space \@p@sheight \space
			\@p@sbbllx \space \@p@sbblly \space
			\@p@sbburx \space \@p@sbbury \space
			startTexFig \space }
	\if@clip{
		\if@verbose{
			\ps@typeout{(clip)}
		}\fi
		\special{ps:: doclip \space }
	}\fi
	\if@angle
		\if@verbose{
			\ps@typeout{(clip)}
		}\fi
		\special {ps:: \@p@sangle \space rotate \space} 
	\fi
	\if@prologfile
	    \special{ps: plotfile \@prologfileval \space } \fi
	\if@decmpr{
		\if@verbose{
			\ps@typeout{psfig: including \@p@sfile.Z \space }
		}\fi
		\special{ps: plotfile "`zcat \@p@sfile.Z" \space }
	}\else{
		\if@verbose{
			\ps@typeout{psfig: including \@p@sfile \space }
		}\fi
		\special{ps: plotfile \@p@sfile \space }
	}\fi
	\if@postlogfile
	    \special{ps: plotfile \@postlogfileval \space } \fi
	\special{ps::[end] endTexFig \space }
}
%
%
\def\psfig#1{\vbox {
	%
	\ps@init@parms
	\parse@ps@parms{#1}
	\compute@sizes
	\ifnum\@p@scost<\@psdraft{
		\PsfigSpecials 
		\vbox to \@p@srheight true sp{
			\hbox to \@p@srwidth true sp{
				\hss
			}
		\vss
                \@AddLabels 
		}
	}\else{
		\if@draftbox{		
			\hbox{\fbox{\vbox to \@p@srheight sp{
			\vss
			\hbox to \@p@srwidth sp{ \hss 
			 \hss }
			\vss
			}}}
		}\else{
			\vbox to \@p@srheight sp{
			\vss
			\hbox to \@p@srwidth sp{\hss}
			\vss
			}
		}\fi

	}\fi
}}

\def\@p@@slabels#1{%
  \def\@p@slabs{#1}%
}

\newcount\@p@sh 
\newcount\@p@sw 
\newcount\@bh
\newcount\@bw
\newcount\@dy
\def\compute@labs{
  \@p@sw=\@p@swidth  \divide\@p@sw by 65536
}%

\newread\labs
\def\@AddLabels{\openin\labs=\@p@slabs\relax
\ifeof\labs\ps@typeout{(no labels)}
\else\closein\labs 
  \ps@typeout{psfig: reading labels from \@p@slabs}
  \compute@labs
  \input \@p@slabs\relax
\fi
}

\newdimen\figsizdim \def\figsiz{\dimen\figsizdim}
\newdimen\fiddle

\newbox\labelbox
\newdimen\xlab \newdimen\ylab
\newdimen\xaux \newdimen\yaux
\newdimen\yy\newdimen\xx
\newdimen\hair\hair=3pt
\def\setlabel#1#2#3#4#5{%
\setbox\labelbox\hbox{$#1$}%
\xlab.5\wd\labelbox 
\xlab#4\xlab%
\fiddle=\hair \fiddle#4\fiddle \advance \xlab by \fiddle
\ylab.5\ht\labelbox 
\ylab#5\ylab %
\fiddle=\hair \fiddle#5\fiddle \advance \ylab by \fiddle
\advance\xlab.5\wd\labelbox%
\advance\ylab.5\ht\labelbox\advance\ylab.5\dp\labelbox%
\vbox to 0pt{\yy=#3 pt 
\divide \yy by \mag%
\multiply \yy by 1010%
\multiply \yy by \the\@p@sw%
\advance\yy by 6pt 
\advance\yy by -\ylab 
\kern-\yy%
\hbox to 0pt{%
\xx=#2 pt 
\divide \xx by \mag%
\multiply \xx by 1010%
\multiply \xx by \@p@sw
\advance\xx by -\xlab
\kern\xx\box\labelbox\hss}\vss}\ifvmode\nointerlineskip\fi}

\def\mathpro{
  \special{ps: plotfile "/home/syer/text/mma.pro-2.1" \space }
}

\psfigRestoreAt
\setDriver

\let\@=\LaTeXAtSign

\newdimen\figsizdim \def\figsiz{\dimen\figsizdim}
\def\calcfigsiz#1{\figsiz=#1 \divide\figsiz by 1010 \multiply\figsiz by \mag}
\def\psfigord#1#2{\psfig{figure=#1.ps,labels=#1.lab,width=#2}}
\def\psfigcall#1#2{\centerline{\psfigord{#1}{#2}}}

\def\psfigodd#1#2#3#4{\vbox{\hbox{\hskip#3{\psfigord{#1}{#2}}}\vskip#4}}

\def\fig[#1,#2,#3,#4]{
  \calcfigsiz{#2} 
  \vbox to #3 {\vss\psfigcall{#1}{\figsiz}}
  \vskip #4
}

\long\def\oddfig[#1,#2,#3,#4,#5,#6]{
        \calcfigsiz{#2}
        \vbox to #3 {\vss\psfigodd{#1}{\figsiz}{#4}{#5}}
        \vskip #6
}

\long\def
\fourfig[#1,#2,#3,#4,#5,#6,#7,#8]{\foursepfig[#1,#2,#3,#4,#5,#6,#7,#8,1]}

\long\def\foursepfig[#1,#2,#3,#4,#5,#6,#7,#8,#9]{
        \calcfigsiz{#5} 
        \vbox to #6 {\vss\centerline
          {\vbox{\vss\psfigodd{#1}{\figsiz}{#7}{0cm}}
           \hskip #9cm
           \vbox{\vss\psfigord{#2}{\figsiz}}}
          \vskip #9cm
          \centerline
          {\vbox{\vss\psfigodd{#3}{\figsiz}{#7}{0cm}}
           \hskip #9cm
           \vbox{\vss\psfigord{#4}{\figsiz}}}
        }
        \vskip #8
}

\long\def\sidefig[#1,#2,#3,#4,#5,#6]{
        \calcfigsiz{#3}
        \vbox to #4 {\eightpoint\vss\centerline
          {\vbox{\vss\psfigodd{#1}{\figsiz}{#5}{0cm}}
           \hskip 0cm
           \vbox{\vss\psfigord{#2}{\figsiz}}}
        }
        \vskip #6
}

\def\offfig{\def\psfigord##1##2{\relax}} 

\def\refit{}\def\refbf{}
\def\ref#1,#2,#3.{{\refit #1\/}, {\refbf #2}, #3\par\noindent}
\def\aa,#1,#2.{\ref{AA},#1,#2.}
\def\acta,#1,#2.{\ref{Acta Astron.},#1,#2.}
\def\annrev,#1,#2.{\ref{ARAA},#1,#2.}
\def\aj,#1,#2.{\ref{AJ},#1,#2.}
\def\apj,#1,#2.{\ref{ApJ},#1,#2.}
\def\apjsupp,#1,#2.{\ref{ApJ\ Supp.},#1,#2.}
\def\apspsci,#1,#2.{\ref{Ap\&SS},#1,#2.}
\def\aasupp,#1,#2.{\ref{AA\ Supp.},#1,#2.}
\def\ica,#1,#2.{\ref{Icarus},#1,#2.}
\def\grg,#1,#2.{\ref{GRG},#1,#2.}
\def\jaa,#1,#2.{\ref{J.\ Astr.\ Astrophys.},#1,#2.}
\def\mnras,#1,#2.{\ref{MNRAS},#1,#2.}
\def\nat,#1,#2.{\ref{Nature},#1,#2.}
\def\pasp,#1,#2.{\ref{PASP},#1,#2.}
\def\pasj,#1,#2.{\ref{PASJ},#1,#2.}
\def\physrev,#1,#2.{\ref{Phys.\ Rev.},#1,#2.}
\def\physrevlett,#1,#2.{\ref{Phys.\ Rev.\ Lett.},#1,#2.}
\def\physrevD,#1,#2.{\ref{Phys.\ Rev.\ D},#1,#2.}
\def\procroy,#1,#2.{\ref{Proc.\ Roy.\ Soc.},#1,#2.}
\def\revmod,#1,#2.{\ref{Rev.\ Mod.\ Phys.},#1,#2.}
\def\sova,#1,#2.{\ref{SvA},#1,#2.}

\def\vec#1{{\bf
\mathchoice
{\hbox{$\displaystyle#1$}}
{\hbox{$\textstyle#1$}}
{\hbox{$\scriptstyle#1$}}
{\hbox{$\scriptscriptstyle#1$}}}}
%

\newcount\footcount \footcount=0
\def\advftncnt{\advance\footcount by1\global\footcount=\footcount}
\def\fonote#1{\relax
  \message{N.B. footnotes do not work}
  \advftncnt\footnote{$^{\the\footcount}$}{#1}
}

\def\etal{{\it et\thinspace al\/}}
\def\adhoc{{\it ad hoc\/}}
\def\ie{{i.e}}
\def\eg{{e.g}}
\def\etc{{etc}}
\def\viz{{\sl viz\/}}
\def\cf{{cf}}

\def\halff{{\textstyle{1\over2}}}
\def\frac(#1/#2){{\textstyle{#1\over#2}}}

\def\reff{\label{1}{ref?}} 

\edef\vbar{|}
\def\subrm#1{_{\rm #1}}
\catcode`|=\active \let|=\subrm

\message{defined subrm}

\def\mod#1{\left\vbar#1\right\vbar} 
\def\inv#1{{1\over#1}} 
\def\pd#1#2{{\upartial#1\over \upartial#2}} \def\lr#1{\left(#1\right)}
\def\td#1{{\rm d}#1}
\def\d#1#2{{\td#1\over\td#2}}
\def\const{\hbox{constant}}
 \psfull

\offlabel

\def\y{{\ifmmode {\rm y} \else y\fi}}

\begintopmatter  

\title{Density cusps: restrictions on non-axi\-symmetric models}
\author{D. Syer} 
\affiliation{ Max-Planck-Institut f\"ur Astrophysik,
85740 Garching-bei-M\"unchen, Germany.}
\author{HongSheng Zhao}
\affiliation{ Sterrewacht Leiden, Niels Bohr Weg 2, Leiden, The Netherlands.}

\shortauthor{D. Syer \& H-S. Zhao}
\shorttitle{Non-axisymmetric cusps}

\acceptedline{}

\abstract{%
Galactic nuclei are now generally thought to have density cusps in
their centres, and the evidence is mounting that as a consequence they
are unlikely to be triaxial.  Self-consistent stellar dynamical models
of non-axi\-symmetric cusps would be an interesting counter-argument
to this conclusion.  We consider 2-d analogues of triaxial cusps: a
sequence of non-axi\-symmetric, cuspy discs first described by Sridhar
\& Touma (1997).  Scale-free models with potential {}$\Phi\propto
r^\alpha$ are examined in detail.  It is shown analytically for
$0<\alpha \la 0.43$ that self-consistent models with positive
phase-space density {\em do not} exist.  Numerical solutions of the
combined Vlasov and Poisson equations suggest that the {\em whole}
sequence of models with $0<\alpha<1$ are also unphysical.  Along with
existing work on cusps, we conclude on purely theoretical grounds that
galactic nuclei are not expected to be triaxial. }

\keywords {galaxies: kinematics and dynamics}

\maketitle

\section{Introduction}
The intrinsic shape of elliptical galaxies is a long-standing problem,
the solution to which may help us to understand the process of galaxy
formation.  Dissipationless collapse and merger models predict that
ellipticals should generally be triaxial with slow figure rotation and
anisotropic velocity ellipsoids (van Albada 1982, Dubinski
\& Carlberg 1991, Navarro, Frenk \& White 1996).  But there is no
strong observational evidence for triaxiality (Merritt 1996).  Results from
deprojection of light---fitting stellar kinematics and statistical analysis
of the flattening vs{.} misalignment---are inconclusive (Franx, Illingworth \&
de Zeeuw 1992, Statler 1994).  In particular, one cannot yet rule out the
possibility that all ellipticals are either prolate or oblate.

One purely theoretical approach to the problem is to examine whether
equilibrium models with strong triaxiality exist.  Recent observations
of the central region of ellipticals have found that the luminosity
density rises as a power-law at small radius (Crane \etal{.} 1993,
Gebhardt \etal{.}  1996).  Schwarzschild (1993) shows that models with
strong cusp and strong triaxiality cannot be in equilibrium for much
longer than the diffusion time of the chaotic orbits.  This is because
the regular box orbits are destroyed by the high density at the
centre.  The boxlets such as banana and fish orbits are not enough to
make a self-consistent model without the chaotic orbits.  Similar
study of 2-d non-axisymmetric models by Kuijken (1993) reaches similar
conclusions.  Merritt \& Fridman (1996) discuss 3-d cuspy models
which are not scale free, and also reach similar conclusions.

\def\ST{Sridhar \& Touma (1997)}
\def\STb{Sridhar \& Touma 1997}
In this paper we examine the existence of cuspy, non-axi\-symmetric
models using a new sequence of St\"ackel models by \ST{}.  These
models in their simplest form are self-similar. They are also
2-dimensional, and as such would not make good models for 3-d galactic
nuclei.  Their interest lies in the fact that they are minimally
symmetric, and hence represent the 2-d analogue of a triaxial system.
Also very important is the fact that they admit only regular orbits,
which makes their interpretation much simpler.  \ST{} postulate that,
since banana orbits are elongated in the direction supporting the
flattened potential, self-consistent realizations of the models could
exist.  If true, this would be a counter example to the results of
previous work on non-axi\-symmetric cusps, and one could argue that
galactic nuclei may indeed be triaxial.

After a brief review of the known results of the models of 
\ST{} in Section 2, we formulate the problem of
testing self-consistency in Section 3.  Analytical results and
numerical results are given in Sections 4 and 5, and Section 6
concludes.

\section{The St\"ackel Potential}
\subsection{Co-ordinates}
It is convenient to switch between two
different coordinate systems.  In the usual polar coordinates
$(r,\theta)$, one can define a new parabolic co-ordinate system
$(\xi,\eta)$,
$$\eqalign{\xi &= r(\cos\theta + 1),\cr \eta &= r(\cos\theta -
1),}\eqno(\n)$$
\na\xietart
so that curves of constant $\xi$ or constant $\eta$ 
are orthogonal parabolas.

\subsection{Potential} 
In the absence of a central black hole, the self-similar
potential of \ST{} is
$$\Phi = \halff r^\alpha[(1+\cos\theta)^{1+\alpha} +
(1-\cos\theta)^{1+\alpha} ],~~~\alpha\in(0,1),\eqno(\n)$$
\na\Phirt
where in their notation we have set $2K=1$ and $\gamma=1-\alpha$.

This potential is in St\"ackel form when written in terms of the
parabolic co-ordinates:
$$\Phi = {\xi^{1+\alpha} + \eta^{1+\alpha}\over\xi-\eta},\eqno(\n)$$
\na\Phixieta
or
$$\Phi={F_+(\xi) - F_-(\eta)\over\xi-\eta}\eqno(\n)$$\na\PhiF where
$$F_+(\xi) = \xi^{1+\alpha}\eqno(\n),$$\na\fpdef
and
$$F_-(\eta)=-F_+(\mod\eta).\eqno(\n)$$\na\fpm

The potential (\Phirt) and the corresponding surface density $\Sigma$
we write as
$$\eqalign{\Phi(r,\theta) &= r^\alpha P(\theta),\cr 
\Sigma(r,\theta) &= r^{\alpha-1} S(\theta).}\eqno(\n)$$
\na\SPdef  
where
$$P(\theta) = \halff [(1+\cos\theta)^{1+\alpha} +
(1-\cos\theta)^{1+\alpha} ].\eqno(\n)$$
\na\Pdef
The functions $S(\theta)$ and $P(\theta)$ are related by Poisson's
equation, which can be written as
$$S(\theta) = \int\limits_0^\infty\!\int\limits_0^{2\pi}\!
\td{w}\td{\phi}\;w^{\alpha-1}
{\alpha^2P(\phi)+P''(\phi)\over[1+w^2-2w\cos(\phi-\theta)]^{1/2}}.\eqno(\n)$$\na\Poiss
where primes represent differentiation (see \STb{}) and we set
$4\pi^2G=1$.

\beginfigure{1} \fig[orbfig,.7\hsize,.7\hsize,1\baselineskip] 
\caption{{\bf Figure \nf.} 
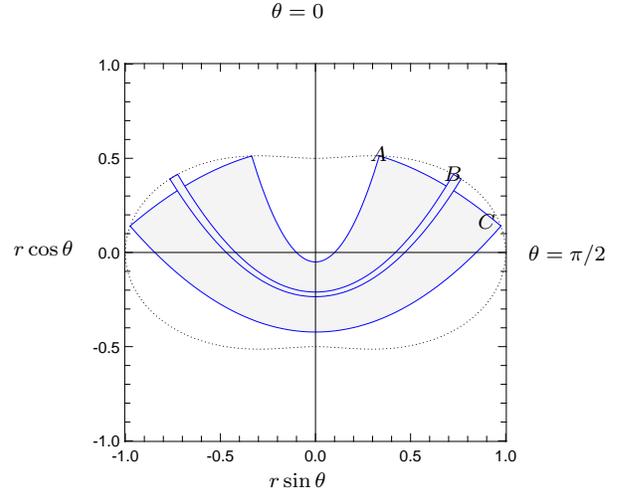
The region covered by two orbits in the model with $\alpha=0.5$.  Both
orbits have $E=1$; the fat one has $I=0.0684$ ($y/y|{max}=0.462$), and
the thin one has $I=0.1478$ ($y/y|{max}=0.998$).  Orbits with negative
$y$ are upside-down version of these ones: mirror images reflected in
$\theta=\pi/2$.  The zero velocity surface $E=\Phi$ is shown as a
dotted line, on which the symbols $A$ and $C$ mark the turning points
of the fat orbit; $B$ marks the ends of the thin orbit.
\naf\orbfig } \endfigure

As a consequence of (\PhiF), the dynamics in the $(r,\theta)$ plane is
analytic.  There are two conserved quantities: the energy $E$ and a
second classical integral $I$, given by
$$\eqalign{I&=2\xi p_\xi^2 - \xi E + F_+(\xi)\cr&=2\eta p_\eta^2 -
\eta E + F_-(\eta),}\eqno(\n)$$ 
\na\Ieq 
where $(p_\xi,p_\eta)$ are the conjugate momenta to $(\xi,\eta)$.  The
orbits of particles are bounded by the co-ordinate lines.  The shapes
of two orbits are shown for illustrative purposes in Figure~\orbfig{}.

\section{Testing self-consistency}
\subsection{Dynamics}
By Jeans'
Theorem, the distribution function is $f(E,I)$, and the associated
surface density can be written
$$\Sigma(r,\theta) = \int f(E,I) \td{^2v}.\eqno(\n)$$\na\Sigdef
The element of phase space in terms of $E$ and $I$ can be derived by
equating $$ \td{p_\xi}\td{p_\eta}\td\xi\td\eta = \td{^2v}\td{^2r}
\eqno(\n)$$
\na\eleq
which is true by virtue of the fact that $(p_\xi,p_\eta)$ are
conjugate momenta.  Hence 
$$ \td{^2v} = {1\over r}
\pd{(\xi,\eta)}{(r,\theta)}\pd{(p_\xi,p_\eta)}{(E,I)}
\td{E}\td{I}.\eqno(\n)$$
\na\jaceq
Now from equation (\xietart) $$\pd{(\xi,\eta)}{(r,\theta)} =
2\sin\theta\eqno(\n)$$
\na\jacxietart and from equation (\Ieq)
$$2r\sin\theta\,\pd{(p_\xi,p_\eta)}{(E,I)} = 
{1\over2\sqrt{(I-I_-)(I_+-I)}}\eqno(\n)$$
\na\jacpsEI
where 
$$\eqalign{I_+ &= -E\eta + F_-(\eta),\cr I_- &= -E\xi +
F_+(\xi).}\eqno(\n)$$
\na\Ipm
Thus
$$2\Sigma(r,\theta) = \int f(E,I)
{\td{E}\td{I}\over\sqrt{(I-I_-)(I_+-I)}},\eqno(\n)$$
\na\SigEI
where the integration is over the range of $(E,I)$ for which the
integrand in real: $I\in[I_-,I_+]$ ($I_+>I_-$ is implied by
$E\ge\Phi$).

\subsection{Self-similarity}
Self-similarity implies that the distribution function can be written
in the form
$$f = E^{-\frac(1/\alpha)} g(IE^{-1-1/\alpha}).\eqno(\n)$$
\na\scaleDF
Thus, defining 
$$Q\equiv{2S\over P^{1-1/\alpha}}.\eqno(\n)$$\na\Qdef 
and setting $z=\Phi/E$ and $y=IE^{-1-1/\alpha}$, the integral
relating $\Sigma$ and $f$ (equation \SigEI) can be written
$$Q(\theta) = \int z^{1/\alpha-2} g(y)
{\td{z}\td{y}\over\sqrt{(y-y_-)(y_+-y)}}\eqno(\n)$$
\na\scaleQz
where $y_\pm$  are defined in the Appendix. 
\beginfigure{2} \fig[regfig,.7\hsize,.7\hsize,1\baselineskip] 
\caption{{\bf Figure \nf.} 
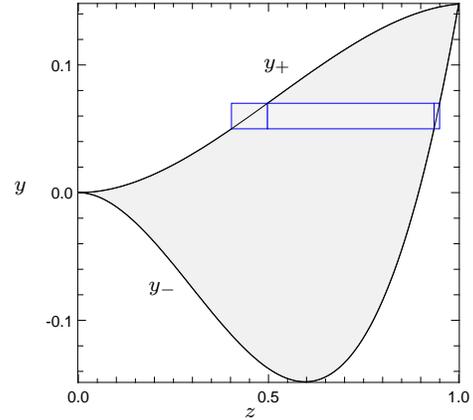
The region of integration in equation (\scaleQz) for $\alpha=0.5$,
$\theta=\pi/3$ is bounded by the curved lines.  For $g(y)$ piecewise
constant it is convenient to split the integral into regions as
shown by the boxes (here $y_j=0.06$ and $\delta_j=0.02$).
\naf\regfig } \endfigure
The region of integration in equation (\scaleQz) is that bounded by the
curves $y=y_\pm(z,\theta)$.  It is illustrated in Figure~\regfig.  

The maximum value of $\mod{y}$ is given by the turning point of $y_+$ with
respect to variations in $z$ and $\theta$.  It is easy to show by
differentiation that the maximum is
$$y|{max}={\alpha\over(1+\alpha)^{1+1/\alpha}}.\eqno(\n)$$\na\ymaxeq
The orbits of particles with $y=\pm y|{max}$ are along the co-ordinate
lines, and hence are thin bananas (\cf{} Bishop 1987)---parabolic thin
shells with their tips lying on the surface $E=\Phi$.  Note that the
largest value of $y$ in the shaded region in Figure~\regfig{} is less
than $y|{max}$, and hence there is no thin banana with $y>0$ for
$\alpha=0.5$, $\theta=\pi/3$.  The smallest value of $y$ is however
$-y|{max}$, so there is a thin banana with $y<0$.

\subsection{Self-consistency}
The problem we would like to solve is to find a distribution
function which self-consistently generates the density and 
the potential in equation (\SPdef);
\ie{.} we seek $g(y)$ such that $S(\theta)$ is given by equations
(\scaleQz) and (\Qdef), with Poisson's equation (\Poiss)
simultaneously satisfied.

Since the model density is symmetric about $\theta=0$ {\em and}
$\pi/2$, we should always weight equally a banana orbit and its
reflection in $\theta=\pi/2$; thus $g(y)$ is even, and we need only
consider positive $y$.  We define $K_+$ as the contribution to $Q$
from a single positive value of $y$:
$$
K_+(\theta,y) = \int z^{1/\alpha-2} 
{\td{z}\over\sqrt{(y-y_-)(y_+-y)}}.\eqno(\n)
$$
\na\Kpint
The corresponding quantity $K_-$ for negative $y$ is just the
reflection of $K_+$ in the major axis, $\theta=\pi/2$:
$$
K_-(\theta,y) = K_+(\pi-\theta,-y).
\eqno(\n)
$$
\na\Kmdef
For each value of $\theta$ there is a contribution to $Q$ from $\pm
y$, and we write
$$
Q(\theta) = \int\limits_{y>0} K(\theta,y)\, g(y) \; \td{y}, \eqno(\n)
$$
\na\Kdef
where
$$
K(\theta,y) = K_+(\theta,y) + K_-(\theta,-y). \eqno(\n)
$$
\na\Kpdef

\beginfigure{3} \fig[kfig,.7\hsize,.7\hsize,1\baselineskip] 
\caption{{\bf Figure \nf.} 
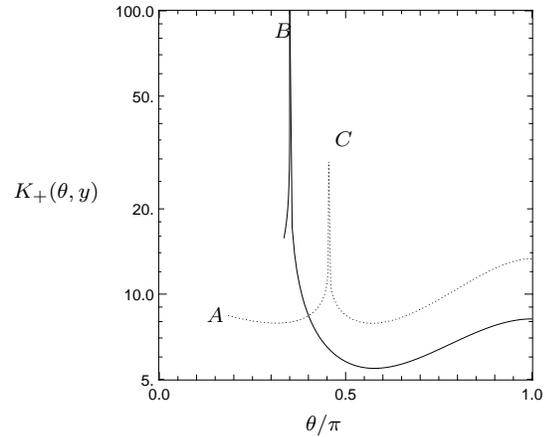
The function $K_+(\theta,y)$ for the two values of $y$ corresponding
to the orbits in Figure~\orbfig{}.  The range of $\theta$ has been
extended to $[0,\pi]$ for clarity.  Thus, by equation (\Kmdef), $K_-$
can be read off from the $\theta>\pi/2$.  The solid curve with the
most pronounced spike is the thin orbit with $y/y|{max}=0.998$; the
dotted curve with a lower, broader spike is the fat orbit with
$y/y|{max}=0.462$.  The symbols $A$ and $C$ mark the turning points of
the fat orbit; $B$ marks the ends of the thin orbit (\cf{.}
Figure~\orbfig{}).
\naf\kfig } \endfigure

The function $K(\theta,y)$ is illustrated in Figure~\kfig{} for two
values of $y$.  The banana shapes corresponding to these orbits are
shown in Figure~\orbfig{}.  The discontinuity and the cusp ($A$ and
$C$) correspond to the the top corner ($A$) and the bottom corner
($C$) of the fat banana in Figure~\orbfig{}.  The end of the thin
banana gives rise to similar features ($B$, \cf{.}  Figure~\orbfig)
which are closer together. Since neither orbit is very sharply bent
(\cf{.} Figure~\orbfig{}), there is no contribution to $K$ at small
$\theta$.  In fact the only orbits which come near $\theta=0$ (the
minor axis) are fat ones---those with small $y$.  The thinner orbit in
Figure~\orbfig{} does not contribute to the density at small $\theta$.
This illustrates that the thin shells cannot be arbitrarily sharply
bent; {}\ie{.} all thin bananas subtend a finite angle.

\subsection{Matrix formulation}
Unfortunately, $y_\pm$ depend on $z$ in such a non-trivial way that
the integral over $z$ in equation (\scaleQz) cannot be simplified 
further analytically.
But the integral over $y$ can be done
analytically for constant $g$, so our approximation will be to make a
grid in $y$ and make $g(y)$ piecewise constant on the grid; thus we
set $g=g_j=\const$ for $y\in(y_j\pm\delta_j/2)$.  We also need a set
of values of $\theta=\theta_i$ at which to try and match the value of
$Q$ with that produced by our guess at $g$.  Thus we recast equation
(\Kdef) as a matrix equation:
$$ Q_i = \sum_j K_{ij} g_j\eqno(\n)$$
\na\mateq
with
$$
K_{ij} =
\int\limits_{y_j-\delta_j/2}^{y_j+\delta_j/2}K(\theta_i,y)\;\td{y}.
\eqno(\n)
$$
\na\Kijdef
We obtain
$$\eqalign{K_{ij} = &\int \td{z}\;z^{1/\alpha-2}\;\times
\cr &\left\{\sin^{-1}h(y_j+\delta_j/2)
- \sin^{-1}h(y_j-\delta_j/2)\right.
 \cr &\left.+ \sin^{-1}h(-y_j+\delta_j/2) -
\sin^{-1}h(-y_j-\delta_j/2) \right\},}
\eqno(\n)$$
\na\KasH
where
$$\eqalign{&1,\qquad y\leq y_+;\cr h(y) = {2(y-y_-)\over
y_+-y_-}\;-\;&1,\qquad y_-<y<y_+;\cr -&1,\qquad y_-\leq y. }\eqno(\n)$$
\na\Hdef
with $y_\pm$ evaluated at $(z,\theta_i)$.  With $h$ defined this way,
the limits of integration in equation (\KasH) can be taken to be
$(0,1)$, but it is more efficient numerically to split the interval
into three parts according to the intersections of the strip
$y\in(y_j\pm\delta_j/2)$ with $y_\pm$. An example is shown in
Figure~\regfig{}: the integration is carried out over each of the
three boxes and the results added together.

To compute $Q(\theta)$ (equation \Qdef), we need to compute the
angular part of the density, $S(\theta)$.  For our purposes it will
suffice to expand $S$ and $P$ in a Fourier series and truncate at high
order:
$$S(\theta)=\sum_{k=0}^n s_k \cos(k\theta)\eqno(\n)$$
\na\sfour
$$P(\theta)=\sum_{k=0}^n p_k \cos(k\theta)\eqno(\n)$$
\na\pfour
where $p_k$ can be evaluated trivially from equations (\Phirt) and
(\SPdef) by standard integrals, and
$$p_k = s_k
B\lr{\frac(k+\alpha+1/2),\halff}B\lr{\frac(k-\alpha/2),\halff}
\eqno(\n)$$
\na\kalnaj
(Kalnajs 1971, Syer and Tremaine 1996).  Since $P(\theta)$ is not only
even, but also symmetric about $\theta=\pi/2$, $p_k$ and $s_k$ vanish
when $k$ is odd.  We truncate at $k=30$, which leads to negligible
error for $\alpha\ge0.5$.

\section{Analytic results}
A necessary condition for self-consistency is that the
orbits can produce the flattening of the model.  An indicator of the
flattening is the ratio of $Q$ on the minor axis to that on the major
axis at the same radius.  Namely
$$
C \equiv {Q(0) \over Q({\pi/2})}.\eqno(\n)
$$
\na\Cdef 
We can also define a similar quantity for each orbit $y$:
$$
C_y\equiv {K(0,y)\over K(\pi/2,y)}.\eqno(\n)
$$
\na\Cydef
To make a self-consistent model, the value of $C$ for the density must
be within the range of the values given by the orbits: \ie{.}
$$
\min[C_y] \le C(\alpha) \le \max[C_y], \eqno(\n)
$$\na\fmvt
for a self-consistent model with a potential cusp $\alpha$. ({\it First
mean value theorem}, inequality 12{.}111, Grad\-shteyn \& Ry\-zhik 1980.)

The maximum of $C_y$ is set by the fattest banana orbit ($y\to0$),
because it spends most of its time near apocentre, near the minor axis
(\cf{.} Figure~\orbfig).  For such orbits, $$C_y\to
\max[C_y]=\cases{\infty,&$\alpha<\halff$,\cr
{\Gamma\lr{\alpha-\halff}\Gamma\lr{\halff+\frac(\alpha/2)}
\over\Gamma\lr{\alpha}\Gamma\lr{\frac(\alpha/2)}}.&$\alpha>\halff$\cr}
\eqno(\n)$$\na\Cmax
The minimum of the $C_y$ occurs for the thinnest
(closed) banana orbit ($y\to y|{max}$).  For such orbit, $C_y=
\min[C_y]=1/\sqrt{2}$. (See the Appendix for derivations of these
limiting values of $C_y$.)

\beginfigure{4} \fig[cafig,.7\hsize,.7\hsize,1\baselineskip] 
\caption{{\bf Figure \nf.} 
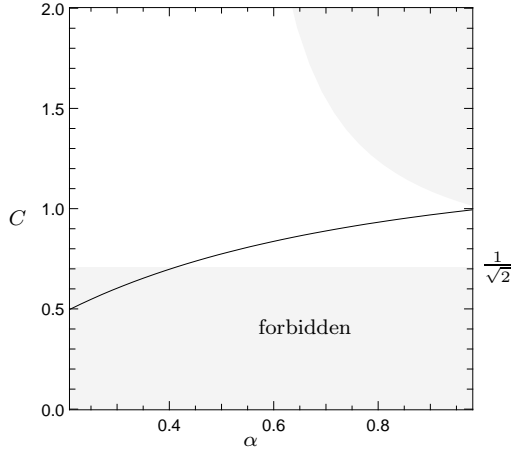
The ratio of the density at the minor axis vs that at the major axis
for the \ST{} models.  The shaded regions indicate those forbidden
from self-consistency by inequality (\fmvt).  Models with
$C<1/\sqrt{2}$ ($0<\alpha\la0.43$) cannot be made self-consistent.
\naf\cafig } \endfigure

Figure~\cafig{} shows $C(\alpha)$ as a function of $\alpha$.  The
inequality
$$
{1 \over \sqrt{2}}\le C(\alpha) \le \max[C_y] \eqno(\n)
$$
\na\caineq
holds only for $\alpha\ga0.43$.  So models with $0<\alpha\la0.43$
cannot be made self-consistent.  The inequality (\fmvt) is only a
necessary condition for self-consistency, so models with higher
$\alpha$ may or may not be self-consistent.  Note, however, that
$\max[C_y]\to1$ {\em and} $C\to1$ as $\alpha\to1$, so (\fmvt) is
only marginally satisfied.  This may be a hint that it will be
difficult to find self-consistent models for $\alpha\simeq1$. To draw
stronger conclusions we must try to solve equation (\mateq)
numerically.

To conclude the section on analytic results, we abandon for a moment
the goal of self-consistency, and remark that when $g(y)=\const$, $Q$
is independent of $\theta$.  Thus $\Sigma\propto \Phi^{1-1/\alpha}$
and the contours of surface density coincide with the contours of the
potential.  As $\alpha$ approaches unity, $\Sigma$ approaches a
constant while $\Phi$ remains highly flattened.  Intriguingly when
$\alpha=1$, $g(y)=\const$ is the trivial self-consistent solution.
This model is however unphysical: $\Sigma$ is constant but it vanishes
for finite $\Phi$.

\beginfigure{5} \fig[chifig,.8\hsize,.8\hsize,1\baselineskip] 
\caption{{\bf Figure \nf.} 
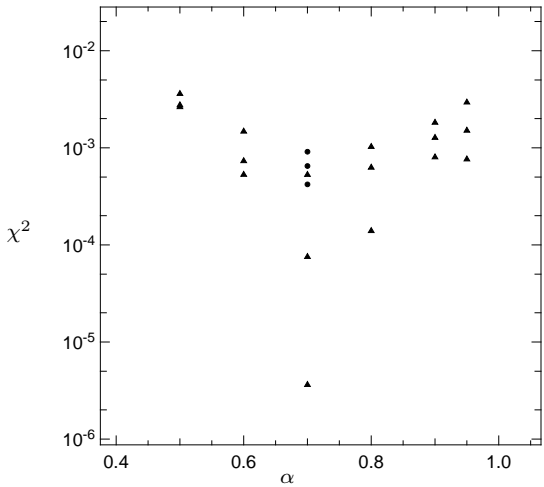
The residuals in the form of $\chi^2$ of best fit orbit models as a
function of $\alpha$.  Triangles are for $N_\theta=11$, and circles
for $N_\theta=20$. {}From top to bottom in each case
$N_y=[25,50,100]$.  All models have significant residual
($\chi^2\gg10^{-12}$).
\naf\chifig } \endfigure
\beginfigure{6} 
\fig[fitifig,.8\hsize,.8\hsize,1\baselineskip] 
\caption{{\bf Figure \nf.} 
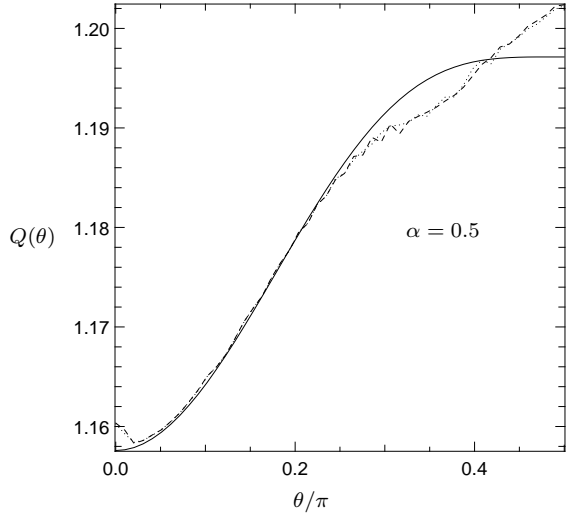
The target density $Q$ (solid line) in the case $\alpha=0.9$.  The
dashed line shows the best fit model with $(N_\theta,N_y)=(50,100)$.
The dotted line shows the best fit model with
$(N_\theta,N_y)=(50,200)$.  The corresponding values of $\chi^2$ are
$0.0026$ and $0.0023$ respectively.  In both cases $\lambda=10^{-12}$,
and consequently $g(y)$ is very far from smooth.
\naf\fitifig } \endfigure

\section{Numerical solutions}
The problem of finding self-consistent solution reduces to solving
equation (\mateq).  A minimum requirement for a physical solution is
that the solution $g(y)$ be positive definite.  It should also be
reasonably smooth.

\def\Smooth{{\cal S}}
Thus we minimize the function
$$\chi^2 = {1\over N_\theta} \sum_{i=1}^{N_\theta} {\left[ \left(\sum
K_{ij}g_j\right) - Q_i\right]^2 \over \sigma^2}
+\lambda\,\Smooth[g],\eqno(\n)
$$\na\chidef and require $$g_j\ge0,\qquad
j\in[1,N_y].\eqno(\n)$$\na\gpos The smoothness is constrained through
$$
\eqalign{\Smooth[g] &= (g_1-g_2)^2 + (g_{N_y}-g_{N_y-1})^2\cr 
& +\;\;\sum_{j=2}^{N_y-1} (2 g_j-g_{j-1}-g_{j+1})^2 }\eqno(\n)
$$
and a small tunable parameter $\lambda$.
We set the characteristic scale for the deviation of the solution from
the target to be
$$
\sigma=Q({\pi \over 2})-Q(0).\eqno(\n)
$$
This problem is solvable with standard routines such as QPROG of IMSL
and E04NAF of NAG. {A source code is in Hanson \& Lawson (1995).  Also
an excellent general discussion of the method is in Press \etal{.}
(1992).}  We use a uniform grid in $\theta$ with
$N_\theta=[11,20,50]$, and a logarithmic grid in $y$ with
$N_y=[25,50,100,200]$.  We apply a minimal smoothing constraint,
$\lambda=10^{-12}$ which leads to solutions with a large amount of
structure on the smallest scales.  Taken literally, this kind of
distribution function is almost certainly unphysical, but it gives the
most generous constraints on self-consistency: if we cannot find even
highly oscillatory solutions, then smooth ones probably do not exist.
We would consider models to be self-consistent if $\chi^2$ was
consistent with zero plus rounding error: \ie{.} $\chi^2
\le 10^{-12}$.   

Figure~\chifig{} shows the value of $\chi^2$ for a range of models.
Figure~\cafig{} suggests that self-consistency should be least
constrained where the solid line is furthest from the forbidden
regions.  Hence we see in Figure~\chifig{} that the smallest value of
$\chi^2$ occurs for $\alpha\simeq0.7$.  Figure~\chifig{} also shows
that additional resolution at $\alpha=0.7$ gives larger values of
$\chi^2$ and thus provides a tighter constraint on self-consistency.
None of the models has $\chi^2$ consistent with zero.

\beginfigure{7} 
\fig[frickefig,.8\hsize,.8\hsize,1\baselineskip] 
\caption{{\bf Figure \nf.} 
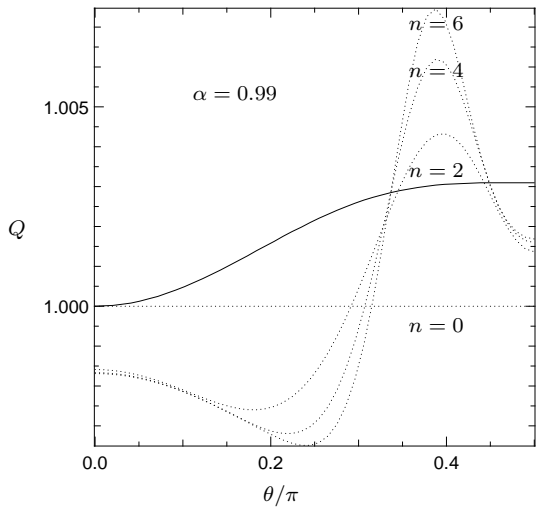
The target density $Q$ (solid line) for $\alpha=0.99$.  Dotted lines
are the contribution to $Q$ from a distribution function with
$g(y)=y^n$, for $n=[0,2,4,6]$.
\naf\frickefig } \endfigure

We note that a more stringent test of self-consistency would perhaps
be the maximum residual $\Delta|{max}$ (as opposed to the mean square
residual $\chi^2$).  From Figure~\fitifig{} we see that typically the
residuals cluster near one or both of the axes ($\theta=0,\pi/2$).
Thus $\Delta|{max}$ can be much greater than $\sqrt{\chi^2}$.

\subsection{Fricke expansion}
To gain some insight to the source of non-self-consistency, we
introduce a basis function approach to solving the Vlasov equation.
Consider the distribution function obtained by setting $g(y)=y^n$ with
$n$ an integer.  This is the equivalent of the method of expansion of
a 3-d axisymmetric system due to Fricke (1951).  Thus we set
$g(y)=y^n$ in equation (\scaleQz) and define
$$Q_n(\theta) = \int z^{1/\alpha-2} y^n
{\td{z}\td{y}\over\sqrt{(y-y_-)(y_+-y)}}.\eqno(\n)$$
\na\Qndef
If $g(y)$ is developed in a Taylor expansion about $y=0$:
$$g(y) = \sum\limits_n\; a_n\, y^n.\eqno(\n)$$
\na\gTaylor
then $Q(\theta)$ is a Fricke expansion:
$$Q(\theta) = \sum\limits_n\; a_n\, Q_n(\theta).\eqno(\n)$$
\na\Qexp
An expression for $Q_n(\theta)$ is derived in the Appendix.
$Q_n(\theta)$ is plotted for $n=[0,2,4,6]$, $\alpha=0.99$ in
Figure~\frickefig{}.  All the $Q_n$ with $n>0$ have broadly similar
shapes.  They are flat for $\theta\ga0$ and have negative curvature;
they rise and reach a maximum at around $\theta=0.4\pi$; and they are
all flat again at $\theta=\pi/2$.  The maximum comes from adding up a
series of cusps like the ones in Figure~\kfig{}.  Its position is
fixed roughly by the minimum angle subtended by the thin bananas.
The shape of the function $Q$, although it is very close to being
constant, is rather different.   We can
understand broadly why the residuals in Figure~\fitifig{} cluster at
the axes.  Given this, it looks rather unlikely
that a Fricke expansion (\Qexp) of $Q$ would converge. 

To find the best fit Fricke expansion for a given model one can
minimize the deviation of equation (\Qexp) from the true value of
$Q(\theta)$ with respect to variations in $a_n$.  In general it is
difficult to control the values of $a_n$ so as to keep $g(y)$ positive
definite.  For this reason we do not recommend Fricke expansion as a
method of solving the Vlasov equation.  

For $\alpha\simeq1$ we know that $g(y)$ is approximately constant, so
there is some hope that the $a_n$ will be small for $n>0$, and hence
that $g(y)$ will remain positive.  We tried fitting $Q(\theta)$ for
$\alpha=0.99$ with a Fricke expansion.  Very poor fits result up to
$n=14$, and are improved very little by going to $n=22$.  The
corresponding $g(y)$ turns out not to be positive definite, and it is
highly oscillatory (more so for the higher order fit, showing the
non-convergent nature of the expansion).

\section{Conclusions}
We attempt to build self-consistent models for the sequence of
St\"ackel potentials of Sridhar \& Touma (1997).  Our numerical
solutions suggest that such models do not exist for the whole range of
$0<\alpha<1$.  We also found a simple analytic argument to show that
models with $0<\alpha \la 0.43$ cannot be self-consistent: all orbits
put too much mass near the minor axis of the potential, and too little
mass near the major axis.  Unfortunately the argument does not apply
to higher $\alpha$.  

We only considered the scale-free models of {}\ST.  There are two
important generalizations which are not scale free: those with central
black holes, and those which are linear superpositions of models with
$\alpha\in(0,1)$.  Consider first a black hole model consisting of a
model with $\alpha\in(0,1)$ but including a black hole of mass
$M$. (The black hole component can be thought of as a model with
$\alpha=-1$.)  Close to the black hole there are lens orbits in
addition to the bananas, and there is a chance that a model composed
largely of lenses could be self-consistent (Sridhar {}\& Touma 1997).
However, at large radius the black hole is insignificant, and the
system should approach the scale-free case.  To be more precise, as
$r\to\infty$ the only orbits which pass close to the black hole will
be increasingly fat bananas ($y\to0$).  In the scale-free case these
orbits have a density biased heavily toward the turning points at the
minor axis, and are therefore unlikely to be useful for building
self-consistent models even in the presence of a black hole.  Thus we
expect self-consistent models to face the same difficulties as in the
present work.  Simple linear superpositions of a finite number of
models with $\alpha\in(0,1)$ will have similar asymptotic problems.

Our results show that regular orbits in a particular set of cuspy
non-axi\-symmetric potentials do not have enough freedom to make a
self-consistent model.  The models we study are 2-dimensional, but
because of their low symmetry they are analogous to triaxial systems.
Had self-consistent models been found it would have been an
interesting counter argument against the conclusions of existing work
on triaxial cusps and galactic nuclei.  In fact, we are in agreeement
with the conclusions of Merritt {}\& Fridman (1996) with respect to a
rather different potential set.  In this case the conclusion was that
the box orbits required to build non-axi\-symmetric models are made
stochastic by the central cusp.  In contrast, the potentials we use do
not admit stochastic orbits, and yet still no self-consistent
equilibria were found.  The existence of triaxial galactic nuclei is
again thrown into doubt by this result.

Acknowledgments: we acknowledge interesting and useful discussions
with Sridhar Seshadri, Tim de Zeeuw and Scott Tremaine.  The authors
are grateful for hospitality provided by their irrespective
institutions and to EARA for facilitating the exchange.

\section*{Appendix}
\subsection*{Definition of $y_\pm$}
The boundary of the region of integration in equation (\scaleQz) is
defined by the quantities
$$\eqalign{y_- &= -uz^{1/\alpha}(1-u^\alpha z),\cr y_+ &=
vz^{1/\alpha}(1-v^\alpha z),}\eqno(\n)$$
\na\ypmz
where
$$\eqalign{(u,v)&=(\xi,\mod\eta)\,\Phi^{-\frac(1/\alpha)}\cr
&=(1+\cos\theta,1-\cos\theta)\;P(\theta)^{-\frac(1/\alpha)}.}\eqno(\n)$$
\na\uvdef
From equations (\Phirt) \& (\SPdef) it follows that
$P(\theta)\ge P(\pi/2)=1$, $u(\theta)\ge u(0)=u(\pi/2)=1$ and
$v\in(0,1)$. 

\subsection*{Limiting values of $C_y$}
Next we derive the value of $C_y$ (equation \Cydef) for the limiting
cases $y\to0$ and $y\to y|{max}$.  

At $\theta=0$ we have $(u,v)=(1,0)$ and $y_+=0$ while $-y_- = x(z)$
where
$$x(z)\equiv z^{1/\alpha}(1-z).\eqno(\n)$$\na\xequiv  Thus $K_+=0$ and
$$\eqalign{K(0,y) &= K_-(0,-y)\cr &= {1\over\sqrt{y}} \int z^{1/\alpha-2}
{\td{z}\over\sqrt{x-y}}.}\eqno(\n)$$
\na\Kzero

At $\theta=\pi/2$ we have $u=v=1$ and $y_+=-y_-=x$.
Thus $K_-=K_+$ and
$$\eqalign{K(\pi/2,y) &= 2K_+(\pi/2,y)\cr &= 2 \int z^{1/\alpha-2}
{\td{z}\over\sqrt{(x-y)(x+y)}}.}\eqno(\n)$$
\na\Kpitwo

As $y\to0$ the numerator in (\Kpitwo) diverges at $z=y$.  Thus the
value of the integral is dominated by the contribution from $z\simeq
0$, and we can replace $x$ by $z^{1/\alpha}$.  Changing variable to
$w=z^{1/\alpha}/y$, we obtain
$$ K(\pi/2,y) \to 2\alpha y^{-\alpha} \int {w^{-\alpha}
\td{w}\over\sqrt{w^2-1}},\eqno(\n)$$\na\Kpz
whence, extending the limits of integration to $[1,\infty)$,
$$ K(\pi/2,y) \to y^{-\alpha} \alpha\sqrt\pi 
{\Gamma\lr{\frac(\alpha/2)}\over\Gamma\lr{\frac(1+\alpha/2)}},
\qquad \hbox{as}\quad y\to0.
\eqno(\n)$$
\na\kplimz
This expression is valid for $\alpha>0$, but for $\alpha\simeq0$ it is
not very accurate.  This can be
rectified by adding the next highest order (logarithmic) term:
$$ K(\pi/2,y) \to \alpha y^{-\alpha} \sqrt\pi 
{\Gamma\lr{\frac(\alpha/2)}\over\Gamma\lr{\frac(1+\alpha/2)}}
+ 2(1+\alpha)\ln{1\over y}.
\eqno(\n)$$
\na\kpliml

For $\alpha<\halff$, as $y\to0$, the integral in equation (\Kzero)
converges, so:
$$K(0,y)\to \sqrt{\pi\over y}\;
{\Gamma\lr{\frac(1/2\alpha)-1}\over\Gamma\lr{\frac(1/2\alpha)-\halff}},
\qquad \hbox{as}\quad y\to0.
\eqno(\n)$$
\na\Kzsmalla
For $\alpha>\halff$, the same trick of replacing $x$ by $z^{1/\alpha}$
works, and we obtain 
$$ K(0,y) \to y^{-\alpha} \alpha\sqrt\pi 
{\Gamma\lr{\alpha-\halff}\over\Gamma\lr{\alpha}}.\eqno(\n)$$
\na\kzlimz

Thus
$$C_y\to\cases{
{\displaystyle y^{\alpha-\halff}\over\displaystyle\alpha}
{\Gamma\lr{\frac(1/2\alpha)-1}\Gamma\lr{\halff+\frac(\alpha/2)}
\over\Gamma\lr{\frac(1/2\alpha)-\halff}\Gamma\lr{\frac(\alpha/2)}},
&$\alpha<\halff$,\cr
{\Gamma\lr{\alpha-\halff}\Gamma\lr{\halff+\frac(\alpha/2)}
\over\Gamma\lr{\alpha}\Gamma\lr{\frac(\alpha/2)}},
&$\alpha>\halff$,\cr}
\eqno(\n)$$\na\Cmaxapp
as $y\to0$.  This is the maximum value of $C_y$.

As $y\to y|{max}$, $x\to y$, and we can set
everything in the integrands for $K_\pm$ constant, except
$\sqrt{x-y}$.  Hence
$$K(0,y)\to {z|{max}^{1/\alpha-2}\over \sqrt{y|{max}}}\int
{\td{z}\over\sqrt{x-y}},\eqno(\n)$$
\na\Kzmax
and
$$K(\pi/2,y)\to \sqrt{2}\;{z|{max}^{1/\alpha-2}\over \sqrt{y|{max}}}\int
{\td{z}\over\sqrt{x-y}}.\eqno(\n)$$
\na\Kpmax
Thus 
$$C_y \to {1\over\sqrt{2}}, \qquad \hbox{as}\quad y\to
y|{max}.\eqno(\n)$$\na\Cmin This is the minimum value of $C_y$.

\subsection*{Coefficients of Fricke expansion}
We derive an expression for $Q_n(\theta)$ (equation \Qexp).  First we
define
$$\eqalign{\bar y&=\halff(y_++y_-)\cr
\Delta&=\halff(y_+-y_-),}\eqno(\n)
$$
\na\ybdef
and then set $$y=\bar y -\Delta \sin \beta\eqno(\n)$$\na\betadef in
equation (\Qndef).  We write
$$Q_n = \sum\limits_{m=0}^{n}\;q_{mn}
\int z^{1/\alpha-2} (y_++y_-)^{n-m}(y_+-y_-)^m\;\td{z}.
\eqno(\n)
$$
\na\Qneq
Since $g(y)$ is even in $y$ we restrict our attention to even $n$.  In
this case the result is
$$
q_{mn}=\cases{2^{1-n}\left({n\atop m}\right)\int\limits_0^{\pi/2}\sin^m\beta
\;\td{\beta},&for even $m$;\cr 0,&otherwise.\cr}
\eqno(\n)
$$
\na\qmneq
Using standard integrals:
$$
q_{mn}=\left({n\atop m}\right){\sqrt\pi\over2^{n}}
{\Gamma\left(\frac(1+m/2)\right)\over\Gamma\left(1+\frac(m/2)\right)}
\eqno(\n)
$$
\na\qmnexp
for even $m$.  In principle the integral in equation (\Qneq) is also
analytic, since $(n,m)$ are integers.

\bye